\documentclass[prx,twocolumn, aps,longbibliography]{revtex4-1}

\usepackage{array}
\usepackage{amsmath,amssymb}
 \usepackage{tikz-feynman} 
\usepackage{graphicx}
\usepackage{subfigure}
\usepackage{mathrsfs}
\usepackage{bm}
\usepackage{mathtools}
\usepackage{epstopdf}
\begin{document}
\title{Landau effective interaction between quasiparticles  in a Bose-Einstein condensate}
\author{A. Camacho-Guardian and Georg M. Bruun } 
\affiliation{ 
Department of Physics and Astronomy, Aarhus University, Ny Munkegade, DK-8000 Aarhus C, Denmark. }
\begin{abstract}
Landau's description of the excitations in a  macroscopic  system in terms of quasiparticles stands out as one of the highlights in quantum physics. 
It provides an accurate description of otherwise prohibitively complex many-body systems, and has led to the development of several key technologies. 
In this paper, we investigate theoretically the Landau effective interaction between quasiparticles, so-called  Bose polarons,
  formed by impurity particles immersed in a Bose-Einstein condensate (BEC). In the limit of weak interactions
between the impurities and the BEC, we derive rigorous results for the effective interaction. They show that it can be strong even for weak impurity-boson interaction, 
if the transferred momentum/energy between the quasiparticles is resonant with a sound mode in the BEC. 
We then develop a diagrammatic scheme to calculate the effective interaction for arbitrary coupling strengths, which recovers the correct weak coupling results. Using this, 
we show that the Landau effective interaction in general is significantly stronger than that  between quasiparticles in a Fermi gas, mainly because a BEC is  more compressible than
a Fermi gas. The interaction is particularly large near the unitarity limit of the impurity-boson scattering, or when the quasiparticle momentum is close to the threshold for 
momentum relaxation in the BEC. Finally, we show how the Landau effective interaction leads to a sizeable shift of the quasiparticle energy with increasing impurity concentration, which should 
be detectable with present day  experimental techniques. 
\end{abstract}
\date{\today}

\maketitle

\section{Introduction}
Landau's quasiparticle theory represents a powerful framework for making accurate  predictions about quantum many-body systems~\cite{Landau1957,Landau1957b}.
 The quasiparticle concept  dramatically reduces the complexity of the problem, and it is exploited across many areas of physics. The quasiparticle framework
 has led to technological breakthroughs such as semiconductor devices and superconductors, and it  is central for the development of  
  future quantum technologies. The effective interaction between quasiparticles plays a key role in Landau's theory,
  where it determines both thermodynamic and dynamical properties.  Interactions between quasiparticles are observed in 
 liquid helium mixtures~\cite{BaymPethick1991book}, and they are the origin of important effects such as conventional and high $T_c$
superconductivity, where the interaction is mediated by 
lattice vibrations  and spin fluctuations respectively~\cite{Schrieffer1983,Scalapino1995}. Likewise, the fundamental 
interaction is caused by the exchange of gauge bosons in particles physics~\cite{Weinberg1995}.

The experimental realisation of highly population imbalanced atomic gases, where the minority (impurity) atoms form quasiparticles, has significantly advanced our 
understanding of this fundamental topic. In particular, since the interaction between the impurity atoms and the majority atoms can be tuned using Feshbach resonances~\cite{Chin2010},
one can study quasiparticle physics systematically and in regimes never realised before. Impurity atoms were first realised experimentally in degenerate 
Fermi gases where they form quasiparticles called Fermi polarons~\cite{Schirotzek2009,Kohstall2012,Koschorreck2012,Scazza2017}. 
We now have an good understanding of the Fermi polaron, even for strong interactions between the impurity and the 
Fermi gas~\cite{Chevy2006,Prokofev2008,Mora2010,Punk2009,Combescot2009,Cui2010,Massignan2011,Massignan2014}. 
Recently,  two experiments  observed long-lived quasiparticles formed by impurity atoms in a Bose-Einstein condensate~\cite{Jorgensen2016,Hu2016}, following 
an earlier experiment on impurities in a one-dimensional (1D) Bose gas~\cite{Catani2012}.
These quasiparticles, called Bose polarons, have been analysed theoretically using a wide range of techniques~\cite{Li2014,Levinsen2015,Shchadilova2016,Tempere2009,Rath2013,Christensen2015,Grusdt2017,Astrakharchik2004,Cucchietti2006,Volosniev2015,Schmidt2016,Ardila2015,Ardila2016}.   

Most of the theoretical and experimental studies  in atomic gases have focused on the properties of a \emph{single} polaron, and 
much less is know about the  interaction between  polarons. In light of its key importance,  a natural question is then whether the great flexibility of atomic gas experiments  
can be used to obtain new insights into Landau's effective interaction, like it did concerning the properties of a single polaron. Studies of 
the Landau effective interaction have been limited to vanishing momenta using perturbation theory~\cite{Yu2012a}, and using variational  and diagrammatic 
techniques for Fermi polarons~\cite{Mora2010,HuM2017,Scazza2017}, or to the case of two fixed impurities in real space using perturbation theory 
and variational techniques~\cite{Dehkharghani2017,Klein2005,Naidon2016}. 

In this paper, we systematically investigate the Landau effective interaction between Bose-polarons, both as a function of momenta and as a function of the 
boson-impurity interaction strength. We derive rigorous results  for weak interaction using perturbation theory. It is shown that even in this limit,  the interaction 
between two polarons can be strong
 when the momentum/energy exchange is resonant with a sound mode in the BEC. Our perturbative calculation is then 
 generalised to arbitrary interaction strengths using a diagrammatic resummation scheme. We show that 
 the Landau effective interaction in general is much stronger than that between Fermi polarons. The interaction is particularly strong when the  boson-impurity interaction 
 is close to the unitarity limit, or when the momentum of one of the polarons approaches the onset of momentum relaxation caused by scattering bosons out of the BEC. 
 The strong effective interaction is then demonstrated to give rise to a substantial shift in the polaron energy as a function of the impurity concentration.
 We conclude by discussing how such effects can be observed experimentally.

\section{Bose polarons} 
We consider mobile impurities of mass $m$, either bosonic or fermionic, immersed in a gas of 
 bosons of mass $m_B$. The Hamiltonian of the system is 
\begin{eqnarray}\label{H1}
H&=&\sum_{\mathbf{k}}\epsilon^B_{\mathbf{k}}b^{\dagger}_{\mathbf{k}}\hat{b}_{\mathbf{k}}+\frac{g_B}{2V}\sum_{\mathbf{k},\mathbf{k}'\mathbf{q}}
{b}_{\mathbf{k}+\mathbf{q}}^\dagger
{b}_{\mathbf{k}'-\mathbf{q}}^\dagger b_{\mathbf{k}'}b_{\mathbf{k}}\\ \nonumber 
&&+\sum_{\mathbf{k}}\xi_{\mathbf{k}}c^{\dagger}_{\mathbf{k}}{c}_{\mathbf{k}}+\frac{g}{V}\sum_{\mathbf{k},\mathbf{k}'\mathbf{q}}{b}_{\mathbf{k}+\mathbf{q}}^\dagger
{c}_{\mathbf{k}'-\mathbf{q}}^\dagger {c}_{\mathbf{k}'}{b}_{\mathbf{k}},
\end{eqnarray} 
where $b^{\dagger}_{\mathbf k}$ and $c^{\dagger}_{\mathbf k}$ create a boson and an impurity, respectively, with momentum $\mathbf k$, and $V$ is the 
volume. Here, 
$\epsilon^B_\mathbf{k}=k^2/2m_B$ is the kinetic energy of the bosons, and $\xi_\mathbf{k}=k^2/2m-\mu$ is the kinetic energy of the impurity atoms relative to their chemical 
potential $\mu$. The boson-boson and boson-fermion interactions are taken to be short range with strengths $g_B$ and $g$, and we assume 
 that there is no direct interaction between the impurities. This is due to the Pauli principle for fermionic impurities, whereas any direct interaction between bosonic impurities 
 can easily be included at the mean-field level.

Below the critical temperature $T_c$, the bosons form a  BEC with  condensate density $n_0(T)$. 
The total density of the bosons is $n_B$, and the density $n$ of the impurities is taken to be so small that the bosons are unaffected.
The BEC  is accurately described by  Bogoliubov theory since we assume that the bosons are weakly interacting with a gas parameter $n_B^{1/3}a_B\ll 1$, where
 $a_B>0$ is the boson-boson scattering length. The normal and anomalous BEC Green's functions for the bosons are  
\begin{align} 
G_{11}(\mathbf{k},z)&=\frac{u_{\mathbf{k}}^2}{z-E_{\mathbf{k}}}-\frac{v_{\mathbf{k}}^2}{z+E_{\mathbf{k}}}\nonumber\\
G_{12}(\mathbf{k},z)&=\frac{u_{\mathbf{k}}v_{\mathbf{k}}}{z+E_{\mathbf{k}}}-\frac{u_{\mathbf{k}}v_{\mathbf{k}}}{z-E_{\mathbf{k}}},
\end{align}
where $E_{\mathbf{k}}=[\epsilon^B_{\mathbf{k}}(\epsilon^B_{\mathbf{k}}+2\mu_B)]^{1/2}$ is the 
Bogoliubov spectrum,   $\mu_B=n_0(T)\mathcal{T}_B$ is the chemical potential of the bosons, and 
$u_{\mathbf{k}}^2/v_{\mathbf{k}}^2=[(\epsilon^B_{\mathbf k+\mu_B})/E_\mathbf{k}\pm 1]/2$ are the usual coherence factors. 
We have defined   $\mathcal{T}_B=4\pi a_{B}/m_B$. 

The impurities interact with the BEC forming quasiparticles  denoted Bose polarons or simply polarons when there is no ambiguity. To avoid confusion, we remind the reader that 
Bose polarons refer to mobile impurities in a Bose gas. The impurities themselves can be either bosonic or fermionic as will indeed be discussed in detail below. 
A polaron with momentum ${\mathbf p}$ is described by the Green's function $G({\mathbf p},z)^{-1}=G_0({\mathbf p},z)^{-1}-\Sigma(\mathbf{p}, z)$, where 
$G_0({\mathbf p},z)^{-1}=z-\xi_{\mathbf p}$ is the non-interacting Green's function and $\Sigma(\mathbf{p},z)$ is the self-energy. 
The polaron energy $\varepsilon_\mathbf{p}$ is found by solving 
 \begin{equation}
\varepsilon_\mathbf{p}=\xi_{\mathbf{p}}+\text{Re}\Sigma(\mathbf{p}, \varepsilon_\mathbf{p}),
\label{Ener}
 \end{equation}
 where we assume that the imaginary part of the self-energy is 
  small and the residue $Z_{\mathbf{p}}=\left(1-\partial \Sigma({\mathbf p},z)/\partial z\right)^{-1}|_{z=  \varepsilon_{{\mathbf p}}}$
  of the quasiparticle is close to unity, so that the polaron is well-defined. 
 In the following, we will for simplicity focus on the case of an attractive interaction  between the impurity and the BEC corresponding to a negative boson-impurity scattering length $a<0$.
  We shall calculate the self-energy $\Sigma(\mathbf{p},z)$ using finite-temperature field theory.

\section{Landau effective interaction} 
Even though there is no direct interaction between the impurities, 
two polarons interact via their effects on the surrounding BEC, or equivalently, due to the exchange of Bogoliubov sound modes.  
Landau's effective interaction between two polarons with momenta $\mathbf{p}_1$ and $\mathbf{p}_2$ is  defined as~\cite{BaymPethick1991book}
\begin{align}
f({\mathbf p}_1,{\mathbf p}_2)=Z_{\mathbf{p}_1}\frac{\delta \varepsilon_{{\mathbf p}_1}}{\delta n_{{\mathbf p}_2}}=Z_{\mathbf{p}_1}
\frac{\delta \text{Re}\Sigma({\mathbf p}_1,\varepsilon_{\mathbf{p}_1})}{\delta n_{{\mathbf p}_2}},
\label{LandauInt}
\end{align}
where $n_{\mathbf p}$ is the quasiparticle distribution function, and we have used (\ref{Ener}) in the second equality. 

\subsection{Weak coupling}\label{weak}
We first consider the weakly interacting regime $k_n |a|\ll 1$, where $k_n=(6\pi^2n_B)^{1/3}$ is the  momentum scale set by the BEC. 
This allows us to  derive  analytical results valid to leading order in $k_na$ and $n_B^{1/3}a_B$, which in  addition to providing important insights, 
 also serve as  a valuable guide for how to construct a strong coupling theory for the effective interaction.

\begin{figure}%[!h]
\begin{center}
\includegraphics[width=\columnwidth]{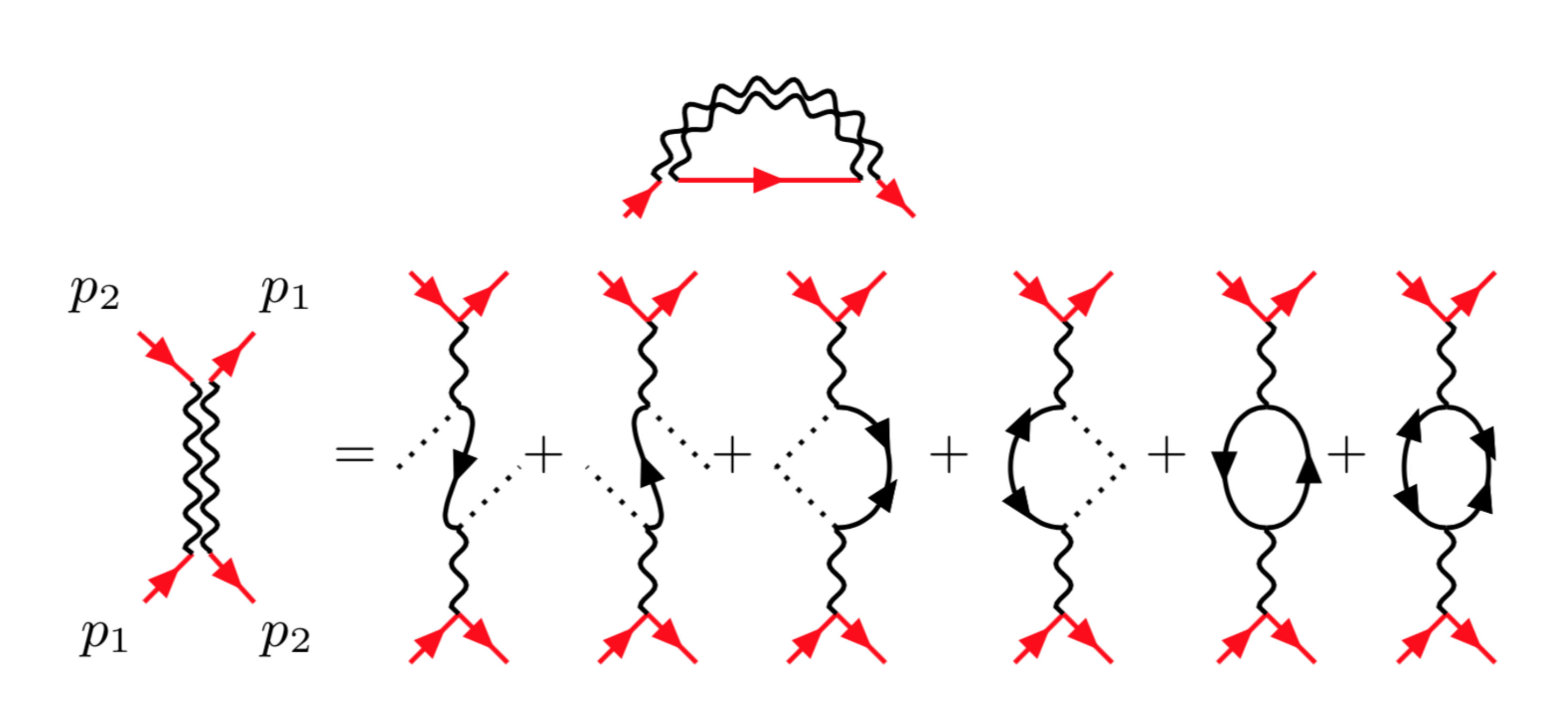}
\end{center}
\caption{Second order self-energy for the Bose polaron as an exchange diagram, where the double wavy line is the second order induced interaction mediated by the BEC. Solid  red lines correspond to the bare Green's function for the impurities, solid black lines are the normal and anomalous Bogoliubov Green's functions for the bosons, dashed lines correspond to condensate particles, wavy lines represent the zero energy boson-impurity vacuum scattering matrix $\mathcal{T}_v$.  }
 \label{Figsecondorder}
 \end{figure}

%%%%%%%%%%%%%%%%%%%%%%%%%%%%%%%%%%%%%%%%%%%%%%%%%

To first order in the scattering length $a$, the polaron energy shift  is  ${\mathcal T}_vn_B$, which is independent of the impurity concentration. It therefore does not 
contribute to the effective interaction. Here ${\mathcal T}_v=2\pi a/m_r$  is the boson-impurity zero energy scattering matrix with $m_r=m_Bm/(m_B+m)$ the reduced mass. 

The second order  contribution to the self-energy $\Sigma_2(\mathbf p,z)$ has been analysed in detail for a single impurity both at zero and non-zero 
temperature~\cite{Christensen2015,Levinsen2017}. Here, we generalise this to a finite impurity concentration $n$. In Fig.\ \ref{Figsecondorder}, we plot the  
second order diagrams for the self-energy as an exchange (Fock) diagram with an induced  interaction. This illustrates 
 that while the induced interaction is inherently attractive in the static case, the sign of the corresponding Landau effective interaction obtained from (\ref{LandauInt})
   depends on the statistics of the impurity particles: It is attractive for bosonic impurities and repulsive for fermionic impurities in the case of a small 
   energy transfer~\cite{Yu2010,Yu2012a}. Note that the contribution to the self-energy 
that is independent of the impurity concentration of course does not depend on the statistics of the impurity particles, since it gives the energy shift of a single polaron~\cite{footnote}.
To second order in $a$, it is enough to 
evaluate the self-energy at the unperturbed quasiparticle energy, 
i.e.\ setting $\varepsilon_\mathbf{p}=\xi_{\mathbf{k}}$ in (\ref{LandauInt})~\cite{Christensen2015}, and to take the residue to be unity. We obtain 
\begin{equation}
f({\mathbf p}_1,{\mathbf p}_2)=\pm\mathcal{T}_v^2\chi(\mathbf{p}_1-\mathbf{p}_2,\xi_{\mathbf{p}_1}-\xi_{\mathbf{p}_2}),
\label{LandauWeak}
\end{equation}
with  $\chi(\mathbf{p},z)=n_0p^2/[m_B(z^2-E^2_p)]$  the density-density correlation function of the BEC at zero temperature.
 Here and in the following  the upper/lower sign  refers to  bosonic/fermionic 
impurities. This effective interaction is independent of the impurity concentration and the mass ratio $m/m_B$ enters only through the 
factor $\mathcal{T}_v^2\propto m_r^{-2}$. Thus, the smaller the reduced mass, the stronger the effective interaction. 
Equation  (\ref{LandauWeak}) recovers  the well-known limit 
\begin{equation}
\lim_{{\mathbf p}_1,{\mathbf p}_2\rightarrow0}f({{\mathbf p}_1,{\mathbf p}_2})=\mp\frac{\mathcal{T}_v^2}{\mathcal{T}_B}=
\mp\left(\frac{\partial \mu}{\partial n_B}\right)^2n_B^2\kappa,
\label{LandauWeakZeroMomentum}
\end{equation}
where $\kappa=n_B^{-2}\partial n_B/\partial \mu_B$ is the compressibility of the BEC at zero temperature. In the second equality, we have used 
$\mu=\mathcal{T}_Bn_B$ to lowest order, so that the Landau interaction for vanishing momenta 
can be written in terms of thermodynamic derivatives~\cite{Viverit2000,Mora2010,Yu2010,Yu2012a}.

In (\ref{LandauWeak}), we have not included the last two "bubble" diagrams in Fig.\ \ref{Figsecondorder}. 
These diagrams are suppressed by a factor $(n_0a_B^3)^{1/2}$ for $T\ll T_c$~\cite{Christensen2015,Levinsen2017} and therefore give a small 
contribution to the effective interaction, except for very low momenta where they give an unphysical divergence for  
$\lim_{{\mathbf p}_1,{\mathbf p}_2\rightarrow0}f({{\mathbf p}_1,{\mathbf p}_2})$ for $T>0$. This comes from an infrared pole in the 
distribution function of the thermally excited bosons $n^B_{\mathbf{k}}=1/[\exp(E_{\mathbf{k}})/T-1]$ as $k\rightarrow 0$.  The interaction with thermally 
excited bosons  is related to the complicated and largely unresolved problem of a systematic theory for a BEC at finite temperature~\cite{Shi1998,Watabe2013}, 
which is beyond the scope of the present paper. As we consider low temperatures here, we neglect these bubble diagrams. 

From a general point of view, we expect the induced interaction to decrease with increasing temperature. The reason is that the Bose gas becomes less compressible as particles are excited out of the BEC~\cite{Fritsch2018}. This effect will however be small for  temperatures 
much smaller than the critical temperature of the BEC, as considered in this paper.

%%%%%%%%%%%%%%%%%%%%%%%%%%%%%%%%%%%%%%%%%%%%%%%%%
\begin{figure}%[!h]
\begin{center}
\includegraphics[width=\columnwidth]{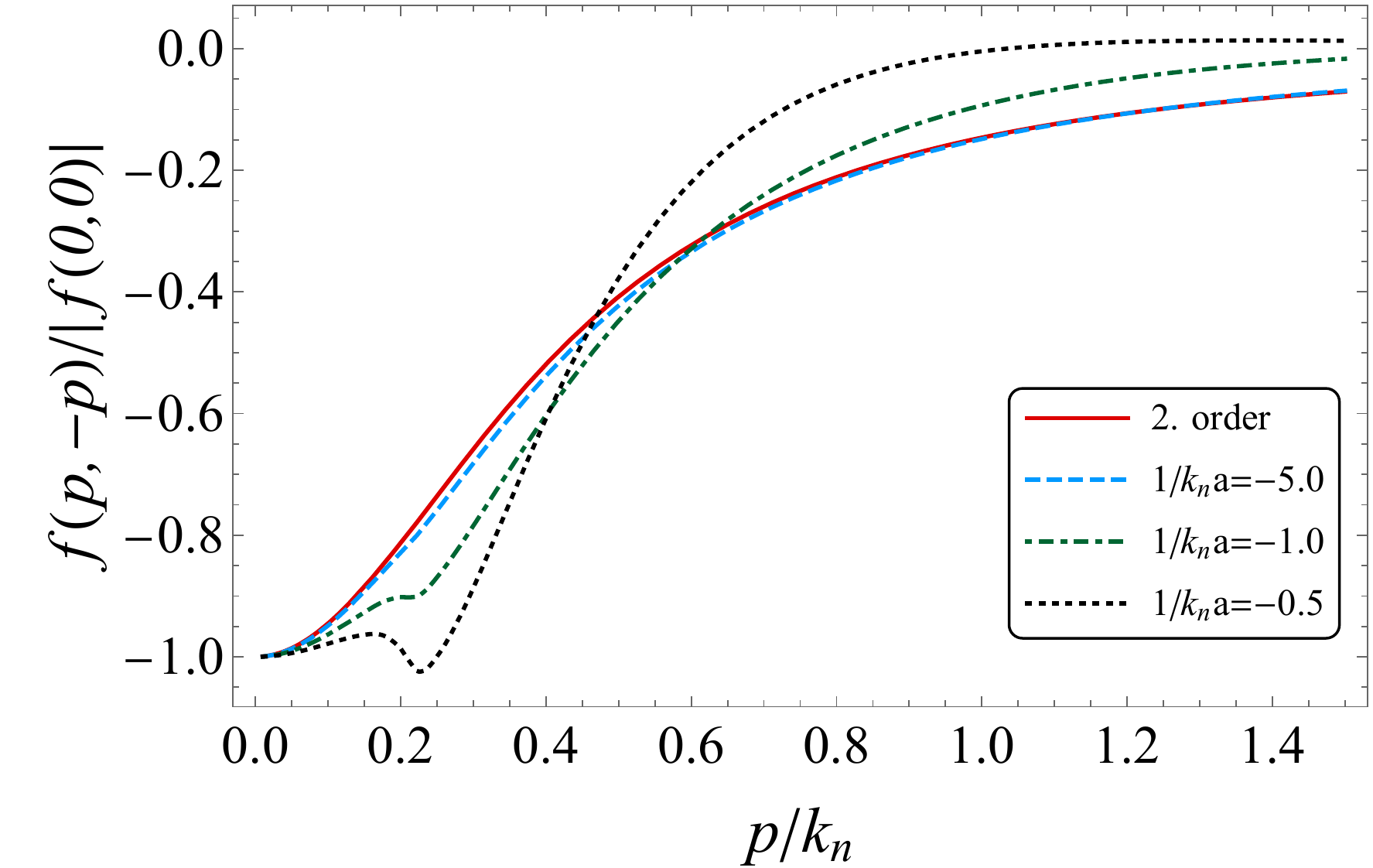}
\end{center}
\caption{The Landau effective interaction $f(\mathbf p,-\mathbf p)$ for zero COM, vanishing impurity concentration, zero temperature,  unit mass ratio $m/m_B=1$, 
and $k_na_B=0.2$. We show the perturbative result (solid red), $1/k_n a=-5.0$ (dashed blue), $1/k_n a=-1.0$ (dot dashed green) and $1/k_n a=-0.5$ (short dashed black). }
\label{Figf(p,-p)}
\end{figure}
%%%%%%%%%%%%%%%%%%%%%%%%%%%%%%%%%%%%%%%%%%%%%%%%%

We plot in Fig.\ \ref{Figf(p,-p)} the zero center-of-mass momentum (COM) Landau effective interaction $f({\mathbf p},-{\mathbf p})$ given by (\ref{LandauWeak}), for zero temperature,
  vanishing impurity concentration, unit mass ratio $m/m_B=1$, and  $k_na_{B}=0.2$. There are no retardation effects for  zero COM and  
  (\ref{LandauWeak}) gives   the well-known Yukawa  interaction. In real space, it is given by 
\begin{align}
f({\mathbf r})=-\frac{\mathcal{T}_v^2n_0m_B}{\pi}\frac{e^{-\sqrt2 r/\xi}}{r},
\label{Yukawa}
\end{align}
where $r$ is the distance between the impurities and $\xi=(8\pi a_Bn_B)^{-1/2}$ is the BEC coherence length. 
We have in Fig.\ \ref{Figf(p,-p)} and (\ref{Yukawa}) taken the impurity to be bosonic, since this is natural for a unit mass ratio $m/m_B=1$. The fermionic case can be obtained by a
simple  sign change. Interestingly, we note that the weak coupling result (\ref{Yukawa}) is not obtained in a variational calculation of the interaction between two fixed impurities~\cite{Naidon2016}.

%%%%%%%%%%%%%%%%%%%%%%%%%%%%%%%%%%%%%%%%%%%%%%%%%
\begin{figure}%[!h]
\begin{center}
\includegraphics[width=\columnwidth]{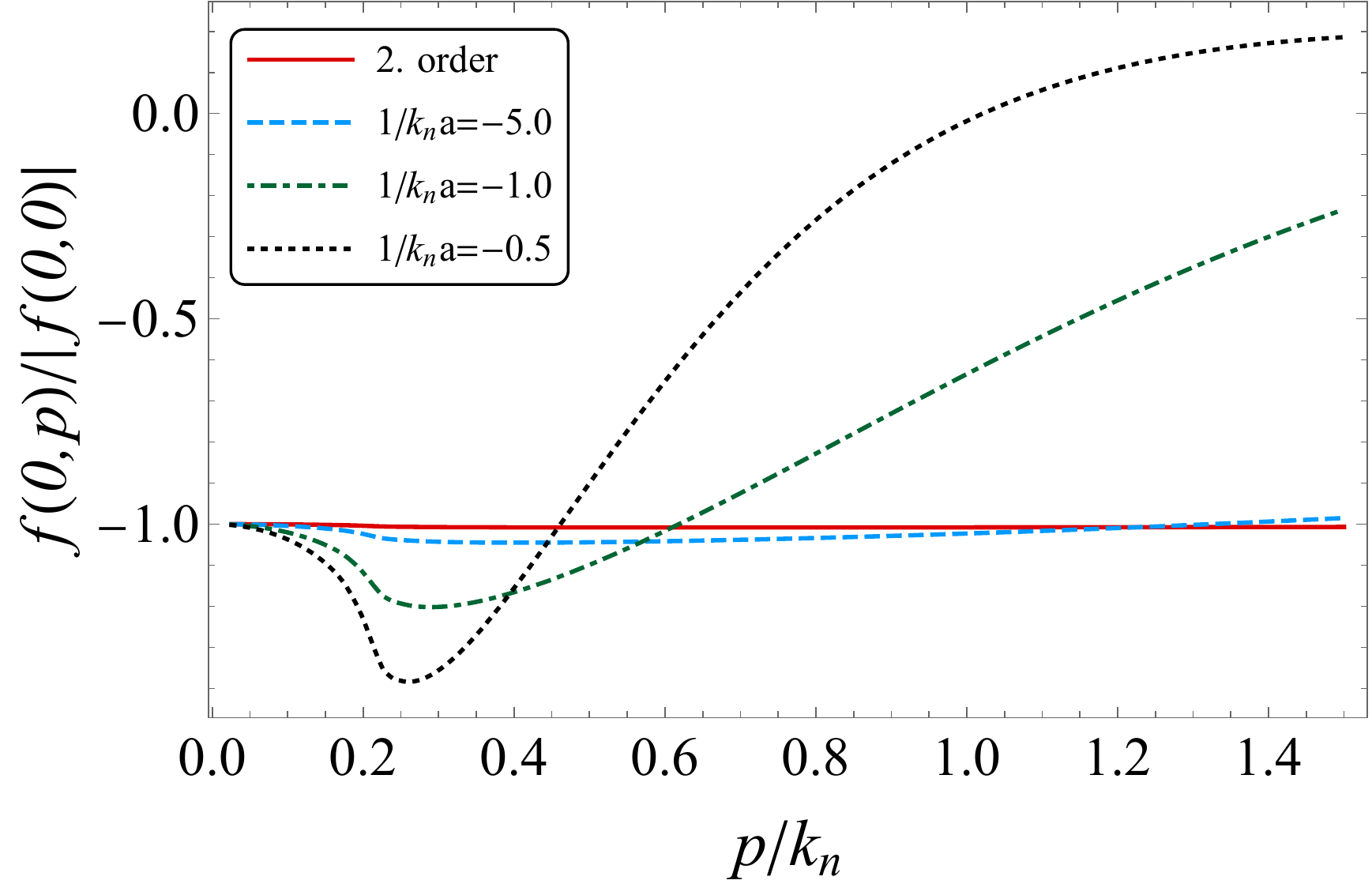}
\end{center}
\caption{The Landau effective interaction $f(0,\mathbf p)$ for unit mass ratio $m/m_B=1$, vanishing impurity concentration, zero temperature, 
$k_na_B=0.2$, and various impurity-boson scattering lengths. The second order perturbative results is also plotted. }
\label{Figf(0,p)alpha1}
\end{figure}
%%%%%%%%%%%%%%%%%%%%%%%%%%%%%%%%%%%%%%%%%%%%%%%%%
%%%%%%%%%%%%%%%%%%%%%%%%%%%%%%%%%%%%%%%%%%%%%%%%%
\begin{figure}%[!h]
\begin{center}
\includegraphics[width=\columnwidth]{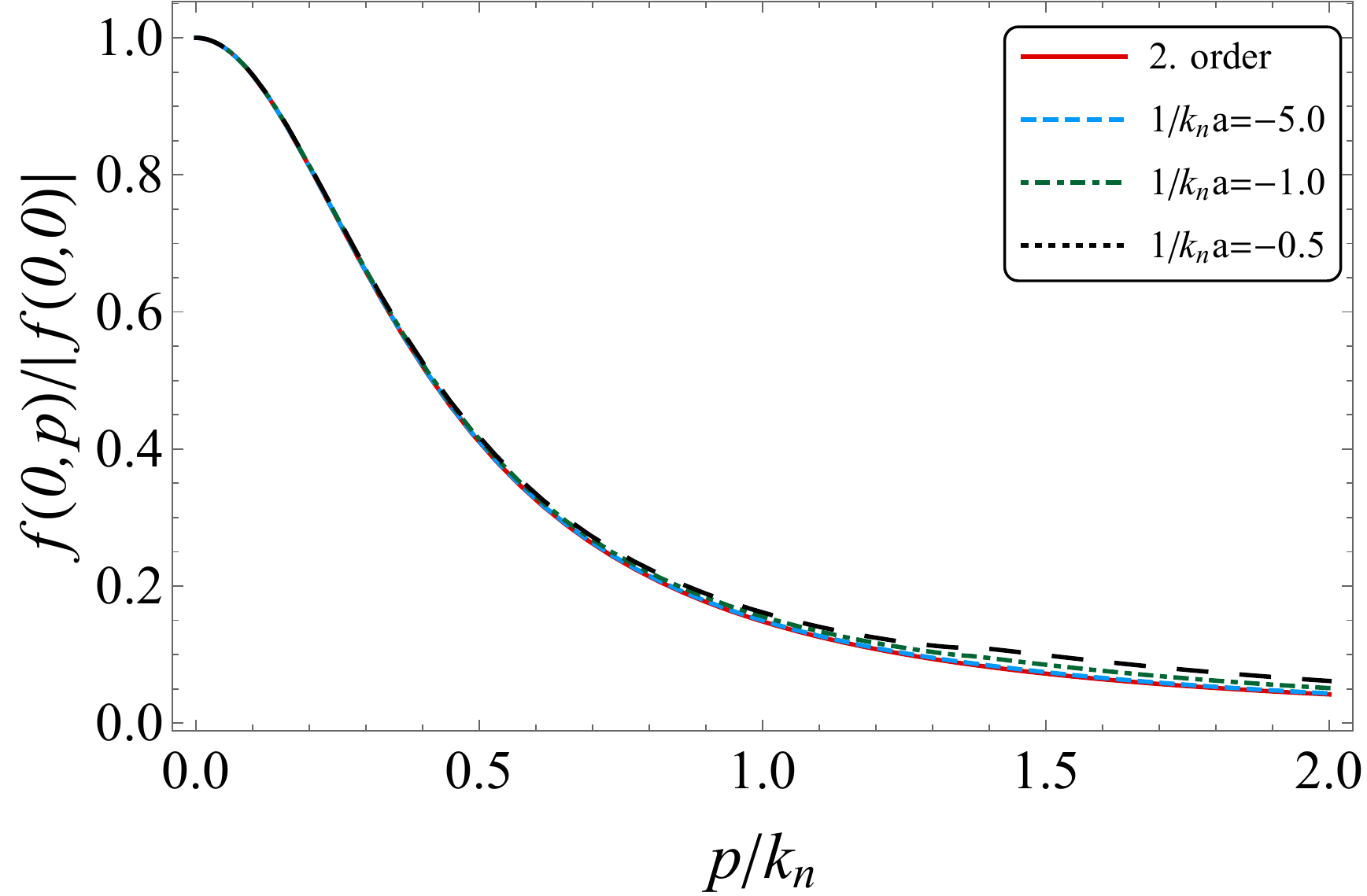}
\end{center}
\caption{The Landau effective interaction $f(0,\mathbf p)$ for  $m/m_B=40/7$, vanishing impurity concentration, $T=0$, 
 $k_na_B=0.2$, and various impurity-boson scattering lengths. The second order perturbative results is also plotted.  }
\label{Figf(0,p)alpha40/7}
\end{figure}
%%%%%%%%%%%%%%%%%%%%%%%%%%%%%%%%%%%%%%%%%%%%%%%%%
%%%%%%%%%%%%%%%%%%%%%%%%%%%%%%%%%%%%%%%%%%%%%%%%%
\begin{figure}%[!h]
\begin{center}
\includegraphics[width=\columnwidth]{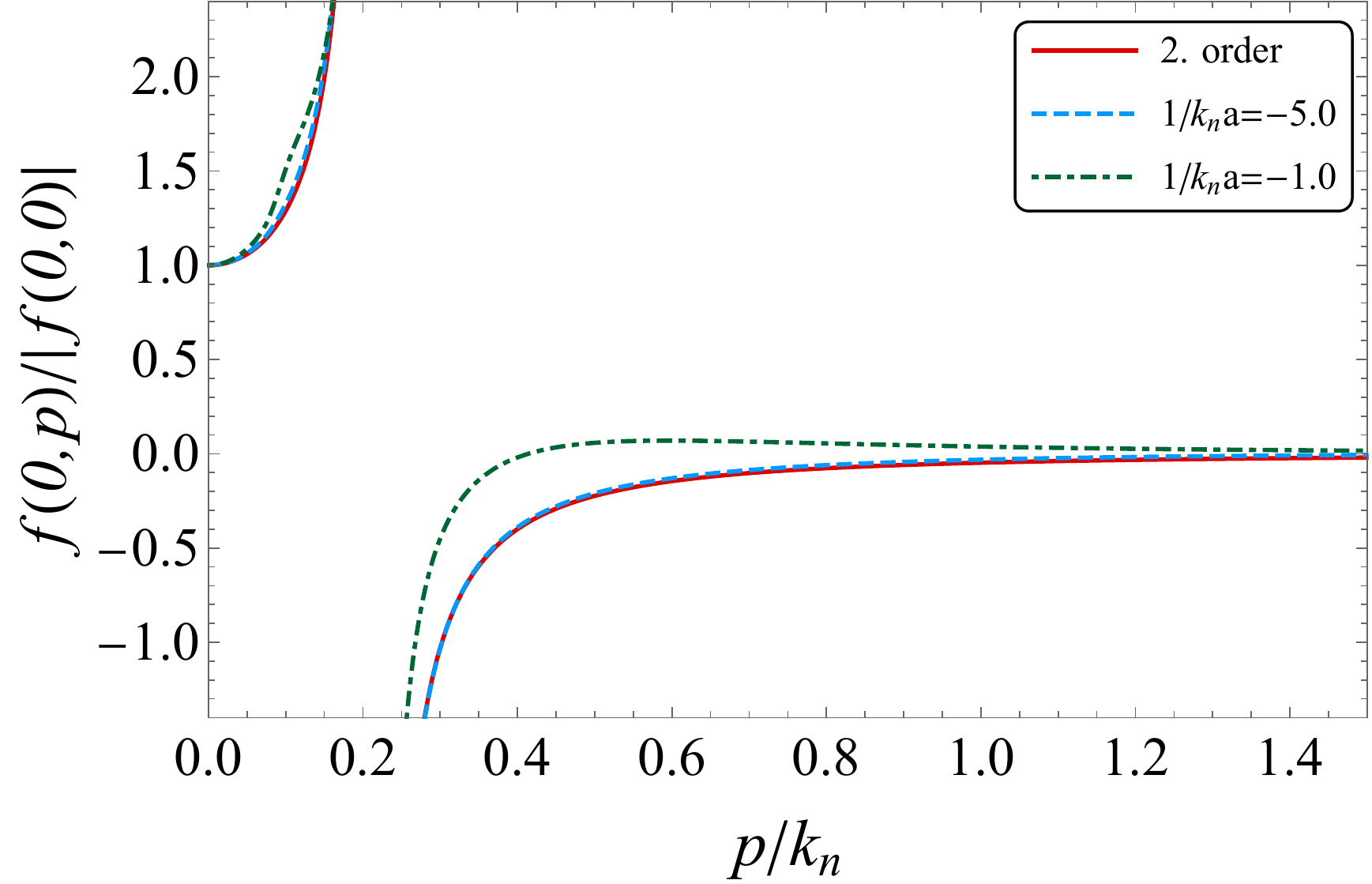}
\end{center}
\caption{The Landau effective interaction $f(0,\mathbf p)$ for  $m/m_B=40/87$, vanishing impurity concentration, $T=0$, 
 $k_na_B=0.2$, and various impurity-boson scattering lengths. The second order perturbative results is also plotted. }
\label{Figf(0,p)alpha40/87}
\end{figure}
%%%%%%%%%%%%%%%%%%%%%%%%%%%%%%%%%%%%%%%%%%%%%%%%%
We plot in  Figs.\ \ref{Figf(0,p)alpha1}-\ref{Figf(0,p)alpha40/87} the Landau effective interaction $f(0,{\mathbf p})$ calculated from (\ref{LandauWeak}), 
which gives the interaction  between a zero momentum polaron and polaron with momentum ${\mathbf p}$ in the weak coupling limit. As above, we take $k_na_B=0.2$, $T=0$, and a vanishing impurity concentration $n$. Figures  \ref{Figf(0,p)alpha1}-\ref{Figf(0,p)alpha40/87} correspond to the mass ratios $m/m_B=1$, 
$m/m_B=40/7$, and $m/m_B=40/87$, which describe the  situation for the  experiments in Aarhus~\cite{Jorgensen2016},  Innsbruck~\cite{Rianne2017}, and JILA~\cite{Hu2016}
respectively. 
 For unit mass ratio, $m/m_B=1$, we have from (\ref{LandauWeak}) $f(0,{\mathbf p})=-\mathcal{T}_v^2/\mathcal{T}_B$ (bosonic impurity), which is  simply a constant 
independent of momentum.  For $m/m_B=40/87$ and $m/m_B=40/7$ on the other hand, there is a strong momentum dependence of $f(0,{\mathbf p})$.  
 The interaction is repulsive for small momenta in both cases since we have taken the  impurities to be fermionic to 
 model the experiments in Refs.~\cite{Hu2016,Rianne2017}. In the case of heavy impurities with $m/m_B=40/7$ shown in Fig.\ \ref{Figf(0,p)alpha40/7}, 
the repulsive interaction decreases monotonically with momentum. 
The  Landau effective  interaction $f(0,{\mathbf p})$ exhibits an interesting behaviour in the
case of light  impurities with $m/m_B=40/87$, as shown in Fig.\ \ref{Figf(0,p)alpha40/87}. It increases strongly with momentum  
 $\mathbf p$ until $p^2/2m=E_p$. Here, it \emph{diverges} even though the boson-impurity interaction is weak. 
The reason is that the transferred momentum/energy $(p,p^2/2m)$ is resonant with a Bogoliubov sound mode in the BEC so that the density-density response 
of the BEC diverges. 
  For larger momentum where $p^2/2m>E_p$,  the energy exchange between the 
polarons is above the Bogoliubov spectrum which means that the density-density response function of the BEC changes sign. The effective interaction between the 
polarons is consequently \emph{attractive} even though the impurities are fermionic.

Note that this resonance effect only happens for light impurities with $m<m_B$
where it is possible to have $p^2/2m=E_p$. In reality, the divergence of the effective interaction when $p^2/2m=E_p$  
will be  softened to a resonance structure for a finite impurity concentration, since the  sound modes of the BEC are damped due to the scattering on the impurities. 
These scattering processes are not included in Bogoliubov theory. For the small impurity concentrations considered in this paper, the damping effects will however
 be negligible, and the broadening of the pole  will be correspondingly narrow.

\subsection{Strong coupling}\label{strong}
A very powerful feature of atomic gases is that one can tune the interaction strength $k_n a$ over orders of magnitude using Feshbach resonances, and we 
 therefore now consider the Landau  effective interaction for arbitrary impurity-boson interaction strength. Here, 
  one has to resort to approximations when calculating the self-energy since there is no small parameter. 
The results obtained above for weak coupling will serve as an important guide to construct a consistent theory. 

\begin{figure}%[!h]
\begin{center}
\includegraphics[width=\columnwidth]{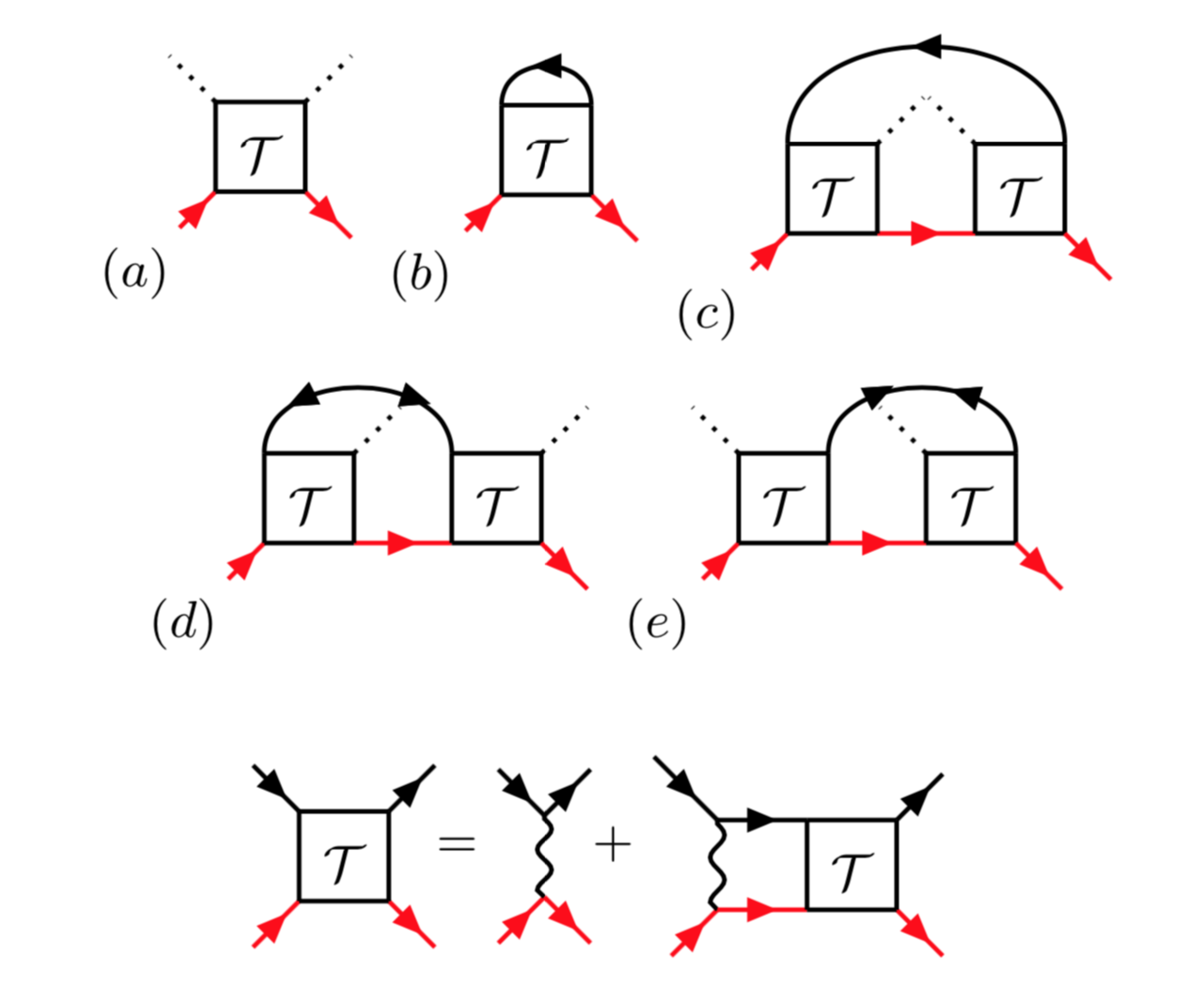}
\end{center}
   \caption{(Top)  Diagrams for the self-energy for a general boson-impurity interaction strength. 
   (Bottom) Ladder approximation for the boson-impurity scattering matrix $\mathcal{T}(p)$.}
   \label{StrongCouplingSelf}
 \end{figure}

\begin{figure}%[!h]
\begin{center}
\includegraphics[width=\columnwidth]{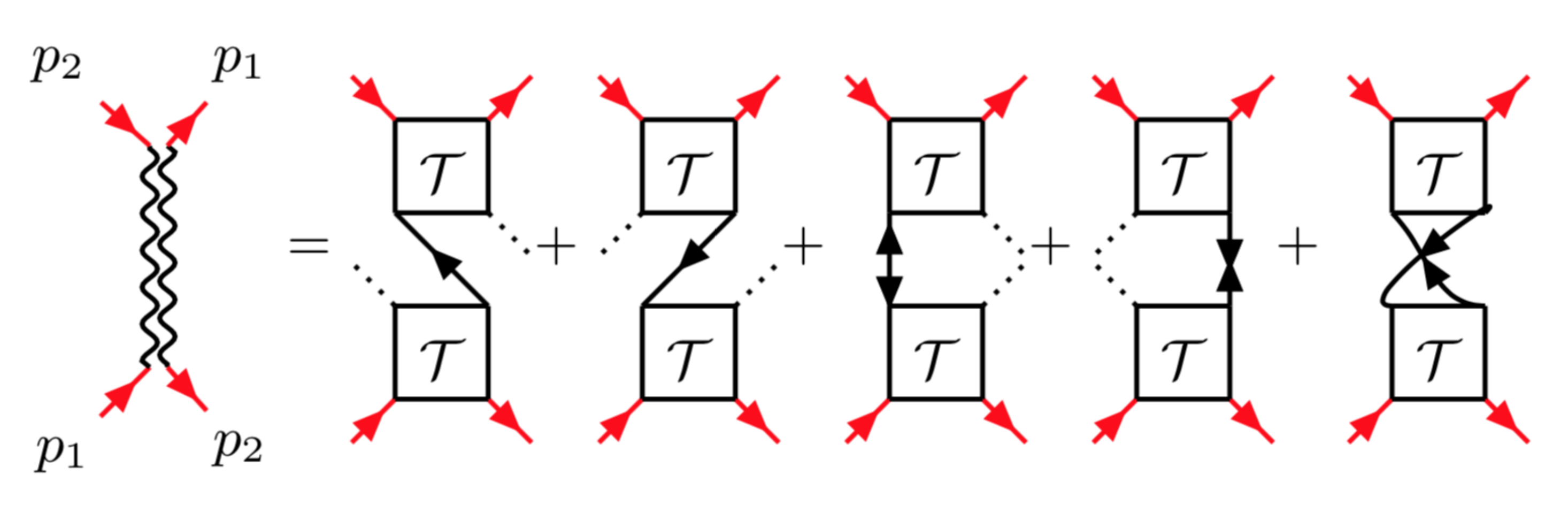}
\end{center}
    \caption{Diagrams for the induced interaction obtained from the self-energy diagrams in Fig.\ \ref{StrongCouplingSelf}. We omit the last diagram in the calculations.}
    \label{StrongCouplingInt}
 \end{figure}

In Fig.\ \ref{StrongCouplingSelf}, we show the  diagrams included in our self-energy calculation, and Fig.\ \ref{StrongCouplingInt} shows the 
corresponding diagrams they generate for the Landau effective interaction. 
These diagrams are obtained by taking the derivative $\delta \Sigma/\delta n_{\mathbf p}$ according to (\ref{LandauInt}), which corresponds to removing an impurity 
line in the self-energy diagrams.
 Diagrams (a)-(b) in Fig.\ \ref{StrongCouplingSelf}  represent the usual ladder approximation, 
which has been extensively used to describe atomic gases. In particular, it has recently been applied to describe the properties of a single Bose polaron~\cite{Rath2013}. 
The ladder approximation is however 
 \emph{not} sufficient to describe the Landau interaction between Bose polarons. This can be seen already in the weak coupling limit, where the ladder approximation only recovers 
 the first diagram in Fig.\ \ref{Figsecondorder}, which gives a \emph{qualitatively wrong} result for the Landau effective interaction. 
Analogously, the ladder diagram in Fig.\ \ref{StrongCouplingSelf} (a) generates only the first diagram in Fig. \ref{StrongCouplingInt}  for strong coupling. 
While this term does recover the microscopic  impurity-boson scattering in the ladder approximation, it fails to  properly describe the propagation 
of density oscillations in the BEC. We  therefore include diagrams (c)-(e) in Fig.\ \ref{StrongCouplingSelf}
 for the self-energy, which generate diagrams the last four diagrams in Fig. \ref{StrongCouplingInt}.
   These diagrams ensure that we describe the density oscillations in the BEC correctly and that we  recover the 
correct weak coupling result.
 
 Similar diagrammatic schemes going beyond the ladder approximation were recently developed 
to analyse Bose-Fermi mixtures~\cite{Guidini2015} and finite temperature effects for  a single polaron~\cite{Guenther2018}. 
The diagrammatic  scheme presented here is however the first to include diagrams (d)-(e) containing anomalous propagators, which is
 crucial for obtaining the correct induced interaction in the perturbative limit, as well as for avoiding unphysical divergencies for strong interactions. 
 The inadequacy of the ladder approximation for obtaining the correct induced interaction 
 should be contrasted to the case of Fermi polarons, where it is sufficient to recover the correct weak coupling result~\cite{Mora2010}.

Two remarks are in order here. First, we do  not include diagram  (b) in Fig.\ \ref{StrongCouplingSelf}, 
which generates the last diagram  in Fig.\ \ref{StrongCouplingInt} for the interaction. The reason 
is the same as for omitting the bubble diagrams  in Fig.\ \ref{Figsecondorder}  for weak coupling. We exclude this diagram because it gives a small contribution to the self-energy and to the 
effective interaction for $T\ll T_c$, except for vanishing momenta where it results in an  unphysical divergence for the effective interaction. 
 Second, the operation $\delta \Sigma/\delta n_{\mathbf p}$ 
generates additional contributions to the effective interaction, which are shown in Fig.\ \ref{StrongCouplingAll} in the Appendix. These terms come from the dependence of the boson-impurity 
scattering matrix $\mathcal T$, given below in (\ref{Tmatrix}), on the impurity concentration $n$. In general, their contribution to the effective interaction is 
small, and we do not include them in the following.

The expression for the self energy corresponding to the diagrams (a) and (c)-(e) in  Fig.\ \ref{StrongCouplingSelf}  is
  \begin{eqnarray}
   \label{Sigma}
  \Sigma(p)&=&n_0\mathcal{T}(p)-n_0\sum_k G_{11}(k)\mathcal{T}^2(k+p)G(k+p)\nonumber\\
    &&-2n_0\mathcal{T}(p)\sum_k G_{12}(k)\mathcal{T}(k+p)G(k+p),
 \end{eqnarray}
 where we have defined the four-momentum 
 $p=(\mathbf p,z)$, and $\sum_k$ is shorthand for $T\sum_{i\omega_n}\int d^3k/(2\pi)^3$ with $\omega_n=2nT$ as Bose Matsubara frequency.
 The boson-impurity scattering matrix is  
\begin{align}
\mathcal{T}(p)=\frac{\mathcal{T}_\nu}{1-\mathcal{T}_{\nu}\Pi_{11}(p)},
\label{Tmatrix}
\end{align}
with $\Pi_{11}(p)=-\sum_kG_{11}(k)G(k+p)$ the in-medium pair propagator.
 In the Appendix, we provide a detailed expression for the self-energy after the Matsubara sum has been performed.

The induced interaction corresponding to the first four diagrams in  Fig.\ \ref{StrongCouplingInt}  is 
 \begin{gather}
 f(\mathbf p_1,\mathbf p_2)= \pm n_0Z_{\mathbf{p}_1}Z_{\mathbf{p}_2}\left[\mathcal{T}^2(p_1)G_{11}(p_1-p_2)+\right.\nonumber\\
 \left.\mathcal{T}^2(p_2)G_{11}(p_2-p_1)+2\mathcal{T}(p_1)\mathcal{T}(p_2)G_{12}(p_2-p_1)\right],
 \label{feff}
 \end{gather}
where $p_i=({\mathbf p}_i,\varepsilon_{\mathbf p_i})$ is the on-shell polaron momentum/energy. In deriving (\ref{feff}),
we have used that the diagrams for $f({\mathbf p}_1,{\mathbf p}_2)$ have to be evaluated with the external outgoing 
momenta ${\mathbf p}_1$ and ${\mathbf p}_2$ swapped 
compared to the ingoing momenta  as indicated in Fig.\ \ref{StrongCouplingInt}, since the Landau interaction is caused by  exchange processes.    
The factor $Z_{\mathbf{p}_2}$ in (\ref{feff}) comes from the residue of the quasiparticle poles of the Green's functions inside the diagrams.

In Fig.\ \ref{Figf(p,-p)}, we plot $f({\mathbf p},-{\mathbf p})$ for  $1/k_na=-5$, 
$1/k_na=-1$, and $1/k_na=-0.5$, and all other parameters as in the perturbative limit discussed in Sec.\ \ref{weak}. We  assume that 
the quasiparticle residues are independent of momentum, i.e.\  $Z_{\mathbf{p}}\simeq Z_0$, so that they cancel out
when normalising by $f(0,0)$. 
The Landau interaction is close to the second order Yukawa result for $1/k_na=-5$. This confirms that our diagrammatic resummation scheme illustrated in 
Figs.\ \ref{StrongCouplingSelf}-\ref{StrongCouplingInt}
recovers the correct  weak coupling limit. On the other hand, we see from Fig.\ \ref{Figf(p,-p)} that the Landau interaction is  
quite different from the Yukawa form for the  stronger coupling strengths $1/k_na=-1$ and   $1/k_na=-0.5$. In particular, it develops 
a minimum around $p\simeq mc$ corresponding to  maximum attraction, where $c=(4\pi n_0a_B)^{1/2}/m_B$ is the speed of sound in the BEC. 
This momentum gives the threshold for momentum 
relaxation at zero temperature: For momenta $p>mc$  the polaron momentum can  decrease by scattering bosons out of the condensate. These processes give rise to an 
imaginary part of the pair propagator $\Pi_{11}(p)$ in the scattering matrix (\ref{Tmatrix}) for $p>mc$ ,  whereas $\Pi_{11}(p)$ is purely real for $p<mc $. 
At the threshold $p=mc$, the  absolute value of $\Pi_{11}(p)$ is minimum. This leads to a large  boson-impurity scattering amplitude and therefore a 
  maximum magnitude of the effective interaction.

  We remark that the physical origin of this maximum is distinct from the origin of the pole  discussed in Sec.\  \ref{weak} in connection with  Fig.\ \ref{Figf(0,p)alpha40/87}. 
  It is purely a strong coupling effect coming from the momentum dependence of the impurity-boson scattering matrix, whereas the pole structure is present even for weak 
  coupling where the scattering matrix is a constant, since it is caused by the transferred energy/momentum being resonant with a Bogoliubov mode.

    We see from \ref{Figf(p,-p)} that the
interaction $f({\mathbf p},-{\mathbf p})$ decreases again for $p>mc$ and it even changes sign
 for large momenta for $1/k_na=-0.5$. This 
is caused by a large imaginary part of the pair propagator $\Pi_{11}(p)$ in the scattering matrix. Physically, it reflects fast momentum relaxation of a polaron with large momentum,
 which however also means that its lifetime  is short making the validity of Landau's quasiparticle description questionable.

We plot in  Fig.\ \ref{Figf(0,p)alpha1} the effective interaction $f(0,{\mathbf p})$ for $1/k_na=-5.0$, $1/k_na=-1$ and $1/k_na=-0.5$, 
and all other parameters as in the perturbative limit.  Again, our diagrammatic  scheme  is close to the  perturbative result 
$f(0,{\mathbf p})=-\mathcal{T}_v^2/\mathcal{T}_B$ for the weak coupling strength  $1/k_na=-5$. The slight momentum dependence for $1/k_na=-5.0$
reflects the weak momentum dependence of the scattering matrix (\ref{Tmatrix}), which is neglected in the perturbative  calculation. 
Again, the Landau effective interaction is very different from the perturbative  result for the stronger coupling strengths $1/k_na=-1$ and $1/k_na=-0.5$. 
We see that it depends strongly on momentum and that there is a minimum, i.e.\ maximum attraction, for $p\simeq mc$. The reason is the same as for 
$f({\mathbf p},-{\mathbf p})$ discussed above: 
The scattering  is resonant when the polaron momentum is at the threshold for 
momentum relaxation. For $p\gtrsim mc$, the strength of the interaction decreases again with momentum  and even changes sign for strong coupling. 
Again, this is however a sign that high momentum polarons  are strongly damped due to momentum relaxation so that they are not  well-defined quasiparticles.  

Figures \ref{Figf(0,p)alpha40/7}-\ref{Figf(0,p)alpha40/87} show $f(0,{\mathbf p})$ for the mass ratios $m/m_B=40/7$ and $m/m_B=40/87$ corresponding to 
fermionic impurities with all other parameters the same as in the perturbative limit.  We see that  the effects of strong interactions are less dramatic for these 
mass ratios. Strong correlations change the quantitative value of the effective Landau interaction but the 
qualitative behaviour is the same as in the perturbative  limit. In particular, the effective interaction  still diverges for light impurities when the transferred momentum/energy is 
resonant with a Bogoliubov sound mode at $p^2/2m=E_p$.

For all mass ratios, $f(0,\mathbf{p})$ and $f(\mathbf{p},-\mathbf{p})$ tend to zero for $|\mathbf{p}|\gg 1/a$ since the scattering becomes suppressed for high momenta.

To illustrate how the strength of the Landau interaction depends on the boson-impurity coupling strength, we plot in Fig.\ \ref{Figf(0,0)} $f(0,0)/Z_0^2$ as a function of $1/k_na$ for 
$T=0$,  $n\rightarrow0$,  $m/m_B=40/7$,
 and three different boson-boson interaction strengths: $k_na_B=0.1$, $k_na_B=0.15$ and $k_na_B=0.2$. From (\ref{feff}), it follows that 
\begin{align}
\frac{f(0,0)}{Z_0^2}=\pm\mathcal{T}^2(0,\varepsilon_{\mathbf p=0})\chi(0,0)=\mp\frac{\mathcal{T}^2(0,\varepsilon_{\mathbf p=0})}{\mathcal{T}_B},
\label{f(0,0)}
 \end{align}
 where the expression $\chi(0,0)$ should be understood as the  limit $\lim_{\mathbf p\rightarrow 0}\chi(\mathbf p,0)$ of the density-density correlation function. 
 Equation (\ref{f(0,0)}) clearly reduces to (\ref{LandauWeakZeroMomentum}) in the weak coupling perturbative regime. 
 From Fig.\ \ref{Figf(0,0)}, we see that $f(0,0)$ increases monotonically with $k_n|a|$ as expected. For weak coupling, we have  $f(0,0)\propto a^2$, whereas it saturates to a maximum value 
   for $k_n|a|\rightarrow\infty$, which depends  both on the mass ratio and $a_B$. 
 %%%%%%%%%%%%%%%%%%%%%%%%%%%%%%%%%%%%%%%%%%%%%%%%%
 \begin{figure}%[!h]
\begin{center}
\includegraphics[width=\columnwidth]{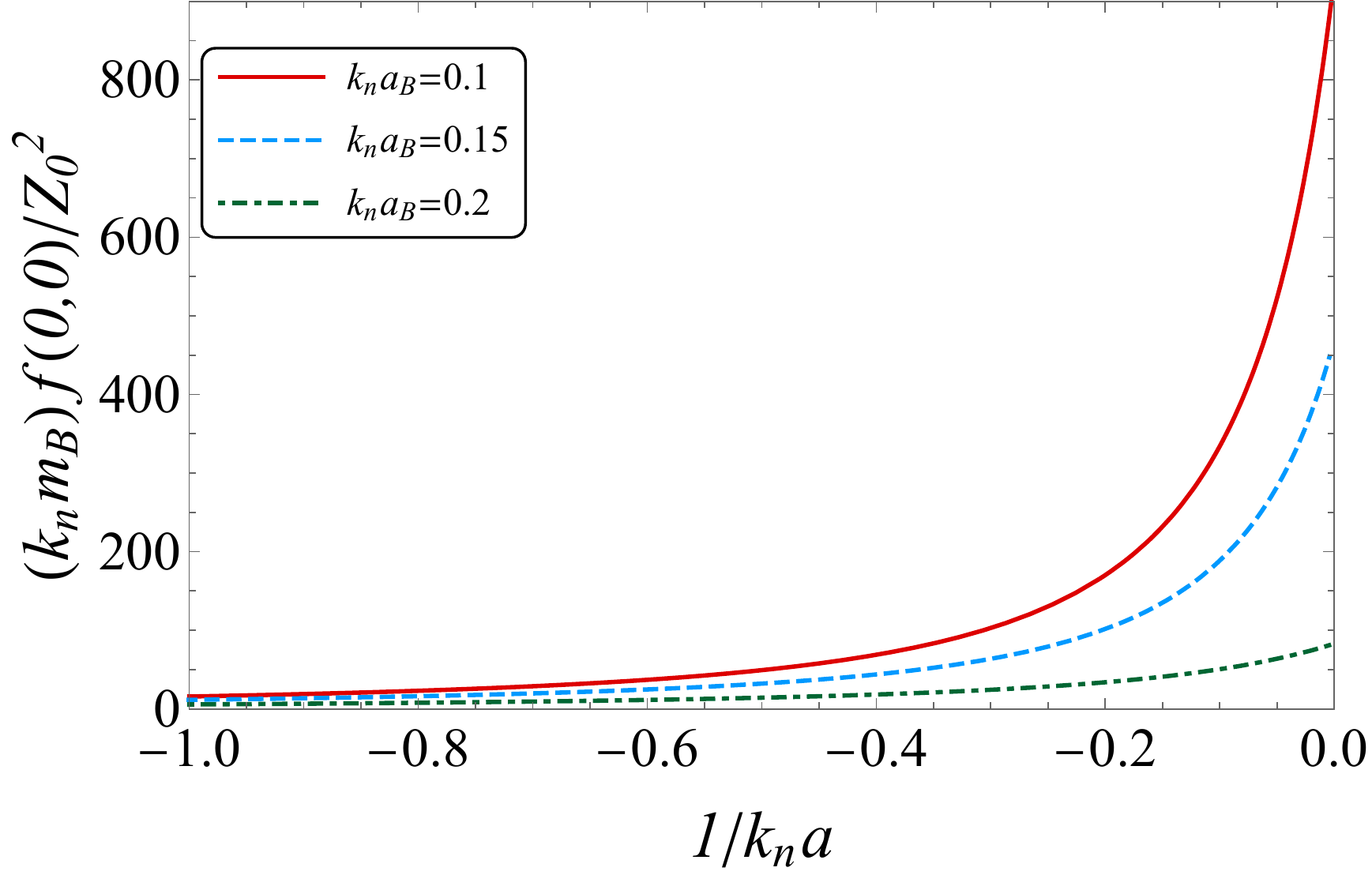}
\end{center}
\caption{The Landau effective interaction $f(0,0)$ as a function of $1/k_n a$ for vanishing impurity concentration, zero temperature,   mass ratio $m/m_B=40/7$, and 
 various boson-boson interaction strengths.}
\label{Figf(0,0)}
\end{figure}
%%%%%%%%%%%%%%%%%%%%%%%%%%%%%%%%%%%%%%%%%%%%%%%%%
This value  is much larger than the corresponding effective interaction between Fermi polarons with vanishing momenta in the unitarity limit. Indeed, 
Monte-Carlo, variational, and ladder calculations yield  $5.0\lesssim mk_Ff(0,0)\lesssim7.1$   for  Fermi polarons
 in the unitarity limit~\cite{Mora2010,Scazza2017,HuM2017}, which is much smaller than what is shown in Fig.\ \ref{Figf(0,0)}.
We note that  $f(0,0)$ remains large even when the factor $Z_0^2$ is included, since the quasiparticle residue for the Bose polaron residue remains 
 significant  for all scattering lengths shown here~\cite{Rath2013,Levinsen2015,Guenther2018}.
  
  The reasons for the larger value of $f(0,0)$ for the Bose polaron as compared to the Fermi polaron are two-fold. First, a weakly interacting
  BEC is much more compressible than the Fermi gas so that 
it mediates density fluctuations more efficiently. This is reflected in the explicit $1/a_B$ dependence in (\ref{f(0,0)}) coming from the fact that the more weakly interacting the BEC, the more compressible it is. 
We also see this in Fig.\ \ref{Figf(0,0)}, which shows that the effective interaction increases with decreasing $k_na_B$. 
Second, both the impurity and the bosons  have vanishing momenta in the scattering processes 
giving rise to the effective interaction $f(0,0)$, since the bosons are scattered out of the condensate. This
 gives rise to the $\mathcal{T}^2(0,\varepsilon_{\mathbf p=0})$ dependence in (\ref{f(0,0)}), which  significantly increases the 
 effective interaction  since the scattering amplitude in general is larger for lower momenta.
In contrast, the impurity-fermion scattering leading to the Landau interaction between Fermi polarons is averaged over all momenta inside the Fermi sea. 

Note however that $f(0,0)$ is not easy to detect directly. As we will  analyse in the next section, observables such as the total energy or the polaron energy 
 involve momentum averages of  $f(\mathbf p_1,\mathbf p_2)$, which  mask a large value of $f(0,0)$. Also, a large value of the effective interaction means that higher 
 order processes not included in this calculation, such as the repeated exchange of phonons in a ladder series  are be important. Such higher order might suppress the 
 magnitude of $f(0,0)$, but it will likely remain  large.

\section{Density dependence of Self-energy}\label{DensityDependence}
The Landau effective interaction $f({\mathbf p}_1,{\mathbf p}_2)$  is  difficult to measure directly. 
For conventional Fermi liquids, momentum averages of $f({\mathbf p}_1,{\mathbf p}_2)$ over the Fermi surface can be extracted from thermodynamic quantities as well as from 
the collective mode spectrum~\cite{BaymPethick1991book}. For atomic gases it has been proposed to use a mixed dimensional Bose-Fermi 
system to detect the induced interaction mediated by a BEC~\cite{Suchet2017}. 
We now discuss how Landau's effective interaction can be probed by measuring the dependence of the polaron energy on the impurity concentration.

It follows from (\ref{LandauInt}) that  
a small change $\delta n_{{\mathbf k}}$ in the polaron distribution function gives rise to the change 
\begin{align}
\delta \varepsilon_{\mathbf{p}}=\int\!\frac{d^3k}{(2\pi)^3}f({{\mathbf p},{\mathbf k}}){\delta n_{{\mathbf k}}},
\label{DeltaELandau}
\end{align}
in the polaron energy. Thus, the dependence of  $\varepsilon_{\mathbf{p}}$  on the impurity density $n$ is a direct consequence of the effective interaction. We write 
the change of a zero momentum polaron energy from its value at zero impurity density as
\begin{align}
\Delta\varepsilon(x)=\varepsilon(x)-\varepsilon(0).
\label{DensityDep}
\end{align}
Here  $x=n/n_B$ is the impurity concentration, and $\varepsilon(x)$ denotes the polaron energy at concentration $x$. We  suppress the momentum label in (\ref{DensityDep})
 and in the following, since we consider a zero momentum polaron from now on.

In the perturbative regime, the induced interaction given by (\ref{LandauWeak}) does not depend on the impurity concentration. 
The change in the ${\mathbf p}=0$ polaron energy due to a non-zero impurity concentration  is then simply 
\begin{align}
\Delta\varepsilon(x)%&=\int\!\frac{d^3k}{(2\pi)^3}f(0,{\mathbf k}) n_{{\mathbf k}}\nonumber\\&
=\pm\mathcal{T}_v^2\int\!\frac{d^3k}{(2\pi)^3} n_{{\mathbf k}}\chi(\mathbf{k},\xi_{\mathbf{k}}+\mu).
\label{DeltaEweak}
\end{align}
Equation (\ref{DeltaEweak}) can  be derived either directly from the second order self-energy or from (\ref{LandauWeak}) and (\ref{DeltaELandau}).
For stronger impurity-boson interaction, the Landau effective interaction does depend on the impurity concentration, and (\ref{DeltaEweak}) no longer holds.

For equal masses with $m=m_B$, (\ref{DeltaEweak}) can easily be evaluated  yielding
$
\Delta\varepsilon(x)=\mp n \mathcal{T}_v^2/\mathcal{T}_B
%\label{ShiftWeakUnitMass}
$. This is a simple consequence of $f(0,{\mathbf p})=\mp\mathcal{T}_v^2/\mathcal{T}_B$ being  constant 
in the perturbative limit for unit mass ratio. The perturbative energy shift (\ref{DeltaEweak}) can also be calculated analytically for any mass ratio $m/m_B$
for fermionic impurities at $T=0$. 
Using  $n_{\mathbf k}=\Theta(k_F-k)$, where $k_F/k_n=x^{1/3}$ is the impurity Fermi momentum, we obtain 
\begin{align}
\label{egalpha}
\frac{\Delta\varepsilon(x)}{E_n}=&(k_n a)^2 \frac{8}{3\pi^2}\frac{(1+\alpha)^2}{(\alpha^2-1)^{3/2}}\Bigg[\frac{k_F}{k_n}\sqrt{\alpha^2-1}-\\ \nonumber
&-\sqrt{\frac{2\alpha^2 m_Bc^2}{E_n}}\arctan\left(\frac{k_F}{k_n}\sqrt{\frac{E_n}{2m_Bc^2}\frac{\alpha^2-1}{\alpha^2}}\right)\Bigg]\\ \nonumber
&\simeq (k_n a)^2\frac{1}{3\pi}\frac{(1+\alpha)^2}{\alpha^2}\frac{x}{k_n a_B}=n\frac{\mathcal{T}_v^2}{\mathcal{T}_B}\frac{1}{E_n},
\end{align}
where $\alpha=m/m_B$ is the mass ratio,  $\sqrt{-1}=i$, and   the last line holds for $x\ll 1$. 
Equation (\ref{egalpha}) diverges  logarithmically for a light impurity with $\alpha<1$,  when 
 the Fermi energy of the impurities is resonant with the Bogoliubov spectrum, i.e.\ when $\epsilon_F= E_{k_F}$. This reflects that the transferred momentum/energy between a polaron at the 
 Fermi surface and the zero momentum polaron is resonant with a sound mode in the BEC. 

This divergence will be softened to a resonance in a real experiment for two reasons. First, any finite temperature will smooth out the Fermi surface. Second, 
 the Bogoliubov modes will be damped due to scattering on the impurities  for a finite impurity concentration, 
 which will broaden the pole of the Landau interaction into a resonance as discussed above. 
 However, for low temperatures and impurity concentrations these effects will be small and the energy shift remains large and 
 non-monotonic as a function of impurity concentration.

\begin{figure}%[!h]
\begin{center}
\includegraphics[width=\columnwidth]{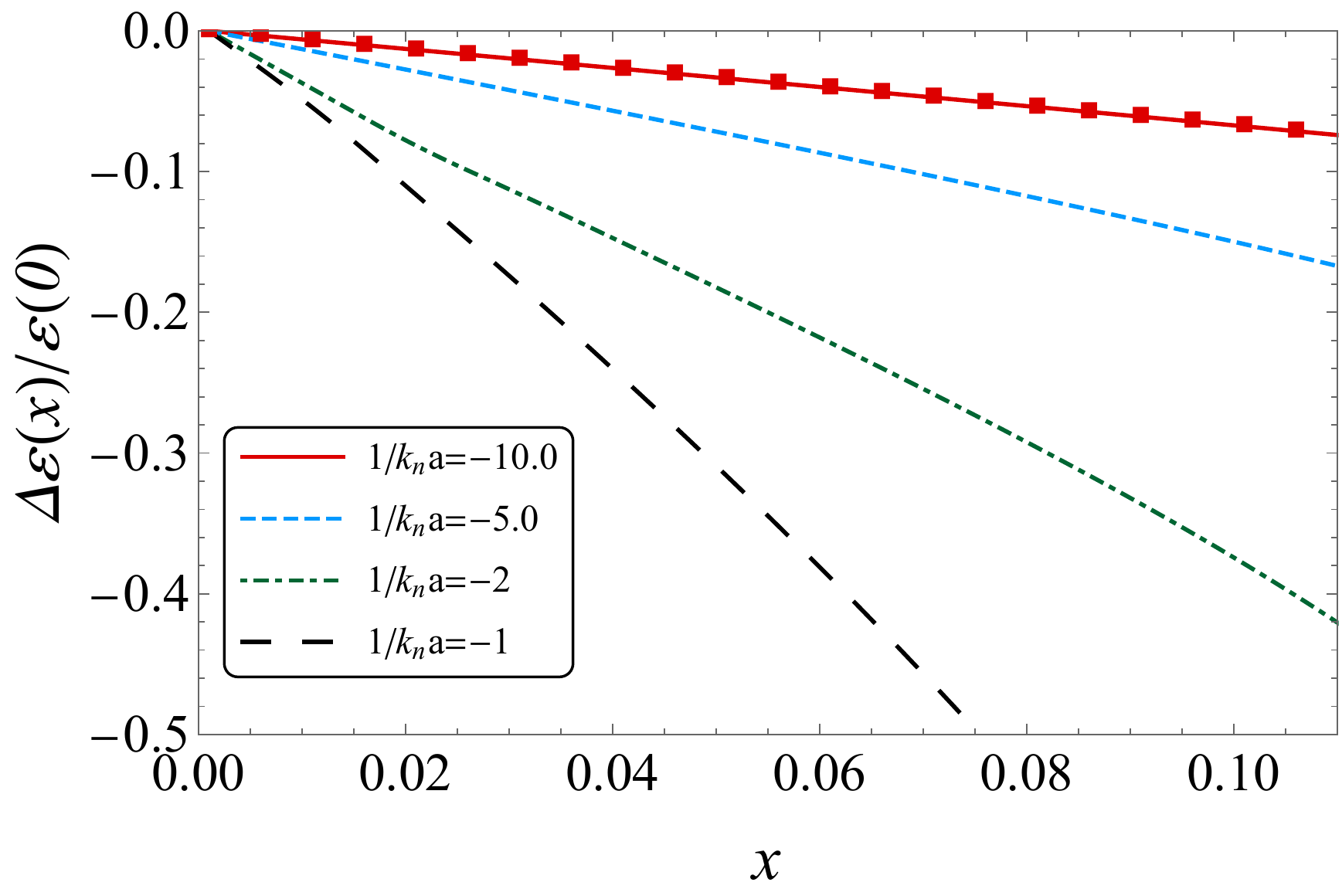}
\end{center}
\caption{The concentration dependence of the  zero momentum polaron energy (\ref{DensityDep}) for $m/m_B=1$, $T=0.4T_c$, $k_na_B=0.2$, and various 
impurity-boson scattering lengths. We also show the perturbative result 
$\Delta\varepsilon(x)=- n \mathcal{T}_v^2/\mathcal{T}_B$ (red squares) for $1/k_na=-10$.}
\label{FigEnergym/mB=1}
\end{figure}

In Fig.\ \ref{FigEnergym/mB=1}, we plot $\Delta\varepsilon(x)$ as a function of impurity concentration $x$ for unit mass ratio $m/m_B=1$,
$k_na_B=0.2$
and different boson-impurity scattering lengths.  We have taken the temperature  $T=0.4T_c$, so that the bosonic impurities remain uncondensed for the range of concentrations shown. 
%%%%%%%%%%%%%%%%%%%%%%%%%%%%%%%%%%%%%%%%%%%%%%%%%
%%%%%%%%%%%%%%%%%%%%%%%%%%%%%%%%%%%%%%%%%%%%%%%%%
 Figure \ref{FigEnergym/mB=1} shows that the polaron energy decreases  with increasing  impurity 
 concentration. This  decrease is caused by a  mainly attractive Landau effective interaction between the bosonic polarons, see Fig.\ \ref{Figf(0,p)alpha1}.  
We see that the concentration dependence of the energy  increases  with the boson-impurity scattering length $a$, even when it is measured in units of the polaron energy at 
zero impurity concentration $\varepsilon(0)$ -- a unit which of course also increases with $a$.
 In the weak coupling limit, this is easily understood from the  fact that the polaron mean-field energy $\mathcal T_vn_B$
scales linearly with $a$, whereas $\Delta\varepsilon(x)=- n \mathcal{T}_v^2/\mathcal{T}_B\propto a^2$. This perturbative result is recovered for 
$1/k_na=-10$ as can be seen in Fig.\ \ref{FigEnergym/mB=1}.
 When impurity-boson interaction is strong, the decrease in the energy is significant. For $1/k_na=-1$, the decrease 
 is around 50\% compared to the polaron energy at zero concentration already at $x\simeq0.075$.

\begin{figure}[!h]
\begin{center}
\includegraphics[width=\columnwidth]{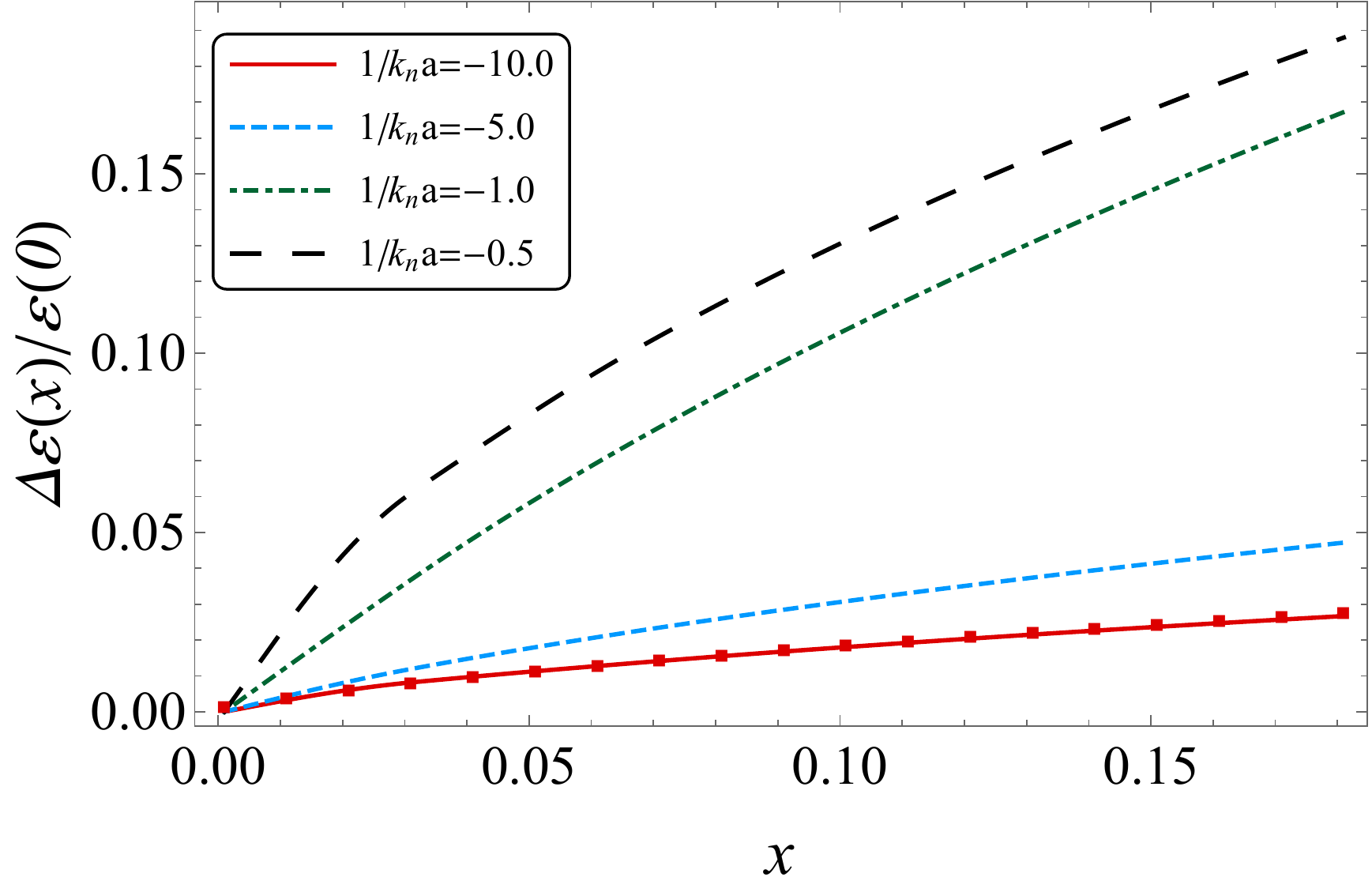}
\end{center}
\caption{The  dependence $\Delta\varepsilon(x)$ of the  zero momentum polaron energy on the impurity concentration $x$ for mass ratio  
$m/m_B=40/7$, $T=0$,  $k_na_B=0.2$, and various 
impurity-boson scattering lengths. We also show the perturbative result (\ref{egalpha}) for $1/k_n a=-10$ (red squares). }
\label{FigEnergym/mB=40/7}
\end{figure}

As an example of fermionic impurities, we plot in Fig.\ \ref{FigEnergym/mB=40/7} the dependence of a zero momentum polaron energy on the impurity concentration 
 for the mass ratio $m/m_B=40/7$. 
%%%%%%%%%%%%%%%%%%%%%%%%%%%%%%%%%%%%%%%%%%%%%%%%%
%%%%%%%%%%%%%%%%%%%%%%%%%%%%%%%%%%%%%%%%%%%%%%%%%
 In contrast to the case of bosonic impurities discussed above, 
the energy now increases with $x$, since the Landau effective interaction is predominantly repulsive between fermionic impurities (see Fig.\ \ref{Figf(0,p)alpha40/7}).
For weak interaction with $1/k_na=-10$, the energy shift is well described by the perturbative result (\ref{egalpha}). For stronger interaction, 
 the concentration dependence of the energy of the fermionic impurities is significant. It is however suppressed compared to the case of bosonic impurities as 
can be seen by comparing   Fig.\ \ref{FigEnergym/mB=40/7} with Fig.\ \ref{FigEnergym/mB=1}. One reason is that Fermi blocking of the impurities 
decreases the effects of the Landau effective interaction. 

\section{Discussion}
Bose polarons have been observed both in the case of bosonic impurities \cite{Jorgensen2016}  and for fermionic impurities \cite{Hu2016}. These experiments 
have focused on the detection of the single quasi-particle properties, namely the energy  and the damping  of the polaron. 
Preliminary data concerning the concentration dependance of the polaron energy was however reported in the Aarhus experiment, which is precisely the 
experimental method for detecting the Landau effective interaction we analyse here. The experiments work with impurity concentrations up to around 10\%, which should give rise to 
observable shifts away from the single impurity limit as can be seen from Figs.\ \ref{FigEnergym/mB=1}-\ref{FigEnergym/mB=40/7}. 
This should be contrasted with the case for Fermi polarons, where the Landau effective interaction  has remained undetected so far in spite of  
concerted experimental efforts~\cite{Scazza2017}. The reason for this difference is of course that the Landau interaction is much larger for Bose polarons than for 
Fermi polarons as we discussed above. We note that for an accurate measurement of the interaction, one must have  
good control of the temperature and impurity concentration. The effects of the trap inhomogeneity should moreover be reduced  or even eliminated. 
We also note that current experimental breakthroughs with ultracold gases offer new exciting mixtures  where both heavy and light impurities can be 
examined~\cite{YuPing2017,Rianne2017, Park2015, Schulze2018, Heo2012}.

Since the effective interaction  is mediated by sound modes in 
the BEC, the corresponding time scale for the effects of polaron-polaron interactions is set by the average distance between the polarons 
divided by the speed of sound in the BEC. In order to observe the Landau effective interaction between polarons, this time scale must be shorter 
than the life-time of the polarons due to for instance  3-body decay.

An important question concerns the accuracy of our theory in the challenging strongly interacting regime $k_n|a|\gtrsim 1$, where there is no small parameter. 
We have very recently   benchmarked our theory
against Monte-Carlo calculations by calculating the binding energy of so-called bi-polarons, which are 
 dimer states of two polarons bound together by the induced interaction. It turns out that the binding energy calculated using the induced interaction obtained with  the theory presented here, 
agrees well with Monte-Carlo results~\cite{Camacho2018}. This shows that our theory is in fact reliable, even  for strong impurity-boson interaction. 
We speculate that a reason for this accuracy,
which is a priori not obvious, is  that it  systematically combines two theories which each have proven to be accurate. The boson-impurity scattering is described 
using the ladder approximation, which has turned out to be surprisingly accurate for cold atomic gases in the strongly interacting regime--
 in particular  for the polaron 
problem~\cite{Kohstall2012,Scazza2017,Chevy2006,Prokofev2008,Mora2010,Punk2009,Combescot2009,Cui2010,Massignan2011,Massignan2014,Hu2016,Rath2013}. 
Moreover, the fact that the  exchange of density oscillations is the  main mechanism leading to an effective interaction in many-body systems
 has been applied successfully in a wide range of physical settings with strong interactions, including liquid Helium and condensed matter systems. 
 We  describe the density oscillations in the weakly interacting BEC using Bogoliubov theory, which is known to be accurate.

Finally, we note that in addition to the dependence of the polaron energy on the impurity concentration, the induced interaction has many other interesting effects.
As mentioned above, it can lead to the formation of dimer states of polarons-- the so-called bi-polarons~\cite{Camacho2018,Devreese2009}. 
The presence of bi-polarons is a many-body effects as they are bound by the induced interaction analyzed in this paper. Bi-polarons therefore become 
unbound with a vanishing BEC density. They  are distinct from 
Efimov states, which are an effect of three-body physics and therefore also stable in a vacuum~\cite{Levinsen2015,Naidon2016}.  On the other hand,  bipolarons can be  stable in the perturbative regime where no Efimov states occur.  Another interesting effect is that since  the induced 
interaction is inherently attractive, it can give rise to pairing between fermionic impurities leading to the formation of unconventional superfluid 
states~\cite{Heiselberg2000,Efremov2002,Illuminati2004,Suzuki2008,Enss2009,Wu2016}.

\section{Conclusions}
We investigated the Landau effective interaction between Bose polarons for arbitrary coupling strengths and momenta. Using perturbation theory, we 
derived analytical results in the limit when the boson-impurity interaction is weak. We 
 showed that for light impurities the interaction can be strong even in the weak coupling regime, 
when the transferred momentum and energy between the polarons is resonant 
with a sound mode in the BEC. To investigate the Landau interaction for arbitrary boson-impurity 
interaction strength, we developed a diagrammatic scheme that recovers the correct weak coupling limit. We showed that the interaction is large when the 
boson-impurity scattering is close to the unitarity limit, or when the momentum of the polaron approaches the threshold for momentum relaxation in the BEC. 
The Landau interaction between Bose polarons is in general much stronger than between Fermi polarons due to the large compressibility of the BEC, and we showed how this 
leads to a substantial shift in the polaron energy as a function of the impurity concentration. We conclude that this shift should be observable using present day experimental technology.

Our results show how the great flexibility of cold atomic gases can be used to explore Landau's theory of quasiparticles systematically and 
in regimes never realised before. Extending the use of this theory is important, given that it forms a powerful platform for our description of many-body systems across a wide range 
of energy scales. Our theoretical scheme, which combines  two theories each known to be accurate, turns out to be reliable even 
in the strongly interacting regime.
It relies on the microscopic  interaction being short range, which is indeed the case in many physical settings where screening effects are significant. 
Our scheme  could therefore  be useful for systems other than atomic 
gases.

\acknowledgements
We acknowledge useful discussions with P.\ Naidon, L.\ Pe\~na Ardila, C.\ J.\ Pethick and T.\ Enss. 
We wish to acknowledge the support of the Villum Foundation.

\begin{widetext}
\appendix
\section{Self Energy and Induced Interaction}
\label{AppendixS}

The pair propagator $\Pi_{11}(p)$ reads as follows,
\begin{align}
\label{ddependance}
\Pi_{11}(p)=\int\! \frac{d^3k}{(2\pi)^3}\left[\frac{u_{\mathbf{k}}^2(1+n^B_{\mathbf{k}}\pm n_{\mathbf{k+p}})}{z-E_{\mathbf{k}}-\xi_{\mathbf{k+p}}}\right.
\left.
+\frac{v_{\mathbf{k}}^2(n^B_{\mathbf{k}}\mp n_{\mathbf{k+p}})}{z+E_{\mathbf{k}}-\xi_{\mathbf{k+p}}} \nonumber +\frac{2m_r}{k^2}\right],
\end{align}
where $n^B_{\mathbf{k}}=1/[\exp(E_{\mathbf{k}})/T-1]$ denotes the bosonic distribution function of the reservoir for $T<T_c$, and 
the last term of $\Pi_{11}(p)$  regularises the pair propagator \cite{Rath2013}.

Performing the Matsubara sums  yields 

\begin{eqnarray}
\label{sigma11}
\Sigma_{11}(p)&=&-\frac{1}{\beta}\sum_{i\omega_\nu}\int \frac{d^3k}{(2\pi)^3}G_{11}(\mathbf{k},i\omega_\nu)n_0\mathcal{T}^{2}(\mathbf{k}+\mathbf{p},i\omega_\nu+z)G(\mathbf{k}+\mathbf{p},i\omega_\nu+z)=\\ \nonumber
&&=n_0\int \frac{d^3k}{(2\pi)^3}\left(G(\mathbf{k}+\mathbf{p},z+E_{\mathbf{k}})u_{\mathbf{k}}^2n^B_{\mathbf{k}}\mathcal{T}^2(\mathbf{k}+\mathbf{p},z+E_{\mathbf{k}})+G(\mathbf{k}+\mathbf{p},z-E_{\mathbf{k}})v_{\mathbf{k}}^2(1+n^B_{\mathbf{k}})\mathcal{T}^2(\mathbf{k}+\mathbf{p},z-E_{\mathbf{k}})\right)+\\ \nonumber
&&\pm n_0\int \frac{d^3k}{(2\pi)^3}n_{{\mathbf{k}+\mathbf{p}}}\text{Re}\left(\mathcal{T}^2(\mathbf{k}+\mathbf{p},\xi_{\mathbf{k}+\mathbf{p}})\right)G_{11}(\mathbf{k},\xi_{\mathbf{k}+\mathbf{p}}-z)\\ \nonumber
&&\mp n_0\int \frac{d^3k}{(2\pi)^3}\int \frac{d\epsilon}{2\pi}n_{\epsilon} \text{Im}\left[2\mathcal{T}^2(\mathbf{k}+\mathbf{p},\epsilon)\right]G(\mathbf{k}+\mathbf{p},\epsilon)G_{11}(\mathbf{k},\epsilon-z).
\end{eqnarray}
for the second term in (\ref{Sigma}), and 
\begin{eqnarray} 
\label{sigma12}
\Sigma_{12}(p)&=&-\frac{2}{\beta}\sum_{i\omega_\nu}\int \frac{d^3k}{(2\pi)^3}n_0\mathcal{T}(p) G_{12}(k)\mathcal{T}(k+p)G(k+p)=\\ \nonumber
&&=-2n_0\mathcal{T}(p)\int \frac{d^3k}{(2\pi)^3}u_{\mathbf{k}}v_{\mathbf{k}}\left(G(\mathbf{k}+\mathbf{p},z+E_{\mathbf{k}})n^B_{\mathbf{k}}\mathcal{T}(\mathbf{k}+\mathbf{p},z+E_{\mathbf{k}})+G(\mathbf{k}+\mathbf{p},z-E_{\mathbf{k}})(1+n^B_{\mathbf{k}})\mathcal{T}(\mathbf{k}+\mathbf{p},z-E_{\mathbf{k}})\right)+\\ \nonumber
&&\pm 2n_0\mathcal{T}(p)\int \frac{d^3k}{(2\pi)^3}n_{\mathbf{k}+\mathbf{p}}\text{Re}\left(\mathcal{T}(\mathbf{k}+\mathbf{p},\xi_{\mathbf{k}+\mathbf{p}})\right)G_{12}(\mathbf{k},\xi_{\mathbf{k}+\mathbf{p}}-z)\\ \nonumber
&&\mp 2n_0\mathcal{T}(p)\int \frac{d^3k}{(2\pi)^3}\int \frac{d\epsilon}{2\pi}n_{\epsilon} \text{Im}\left[2\mathcal{T}(\mathbf{k}+\mathbf{p},\epsilon)\right]G(\mathbf{k}+\mathbf{p},\epsilon)
G_{12}(\mathbf{k},\epsilon-z).
\end{eqnarray}
\end{widetext}
for the third term in (\ref{Sigma}). We have defined $n_{\mathbf{k}}=(\exp\beta\xi_{\mathbf k}\mp1)^{-1}$ and $n_{\epsilon}=(\exp\beta\epsilon\mp1)^{-1}$ where the upper/lower sign as usual 
is for bosonic/fermionic impurities. 

\begin{figure}%[!h]
\begin{center}
\includegraphics[width=\columnwidth]{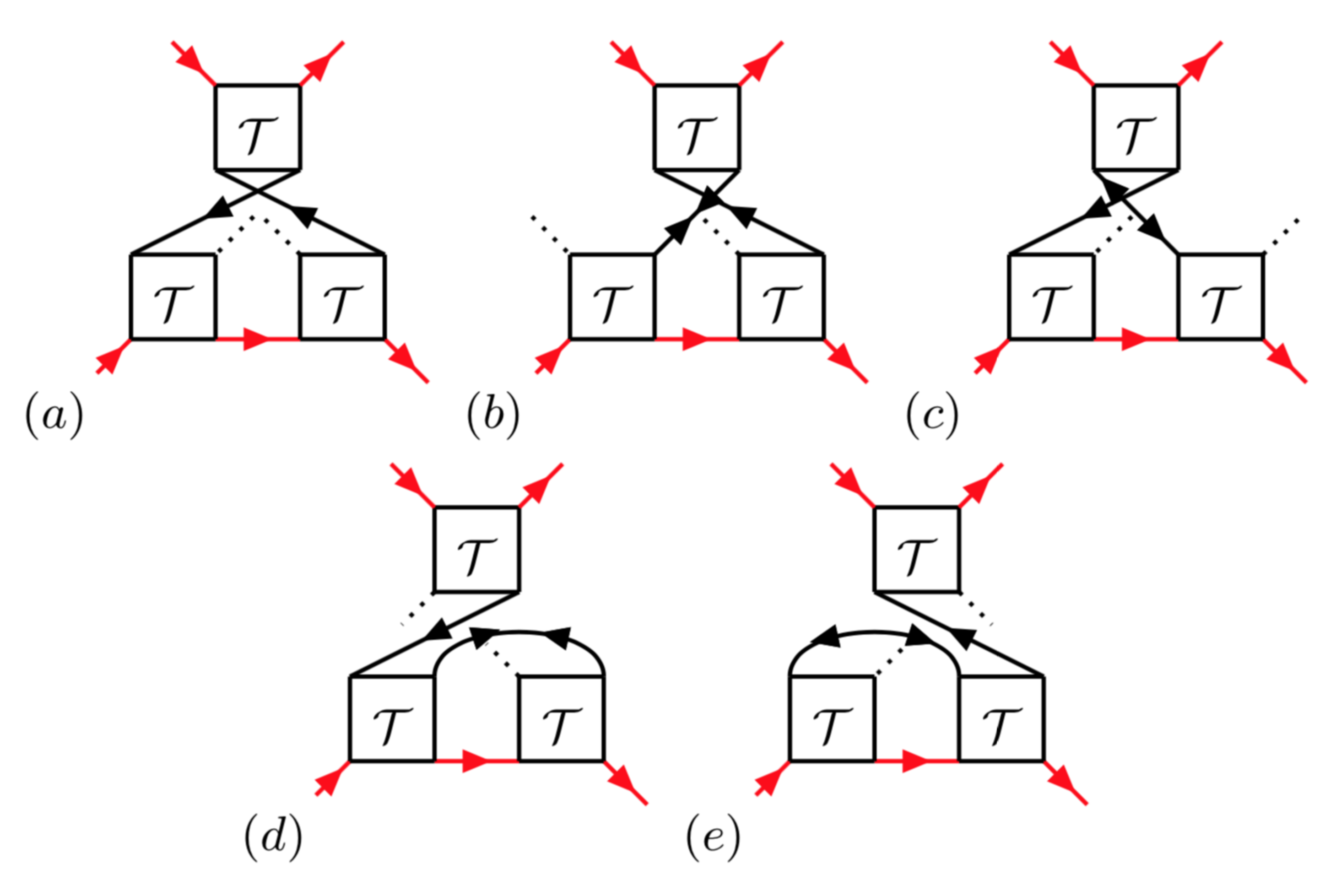}
\end{center}
  \caption{Diagrams obtained from the dependence of the ${\mathcal T}$-matrices in Fig.\ \ref{StrongCouplingSelf} on the impurity density. We do not show the 
  equivalent diagrams, where two ${\mathcal T}$-matrices are on top and one below. 
   }
  \label{StrongCouplingAll}
 \end{figure}

The diagrams for the effective interaction coming from the dependence of the ${\mathcal T}$-matrices  in Fig.\ \ref{StrongCouplingSelf} on the impurity concentration  are shown in 
Fig.\  \ref{StrongCouplingAll}. They can be obtained by removing an impurity line inside one of the ${\mathcal T}$-matrices, 
which produces diagrams for the interaction with three ${\mathcal T}$-matrices. We do not include the diagrams in Fig.\  \ref{StrongCouplingAll} when 
calculating the effective interaction, since they do not represent  a consistent inclusion of all diagrams with three ${\mathcal T}$-matrices.

\bibliography{PolaronsInduced1} 

\begin{thebibliography}{63}%
\makeatletter
\providecommand \@ifxundefined [1]{%
 \@ifx{#1\undefined}
}%
\providecommand \@ifnum [1]{%
 \ifnum #1\expandafter \@firstoftwo
 \else \expandafter \@secondoftwo
 \fi
}%
\providecommand \@ifx [1]{%
 \ifx #1\expandafter \@firstoftwo
 \else \expandafter \@secondoftwo
 \fi
}%
\providecommand \natexlab [1]{#1}%
\providecommand \enquote  [1]{``#1''}%
\providecommand \bibnamefont  [1]{#1}%
\providecommand \bibfnamefont [1]{#1}%
\providecommand \citenamefont [1]{#1}%
\providecommand \href@noop [0]{\@secondoftwo}%
\providecommand \href [0]{\begingroup \@sanitize@url \@href}%
\providecommand \@href[1]{\@@startlink{#1}\@@href}%
\providecommand \@@href[1]{\endgroup#1\@@endlink}%
\providecommand \@sanitize@url [0]{\catcode `\\12\catcode `\$12\catcode
  `\&12\catcode `\#12\catcode `\^12\catcode `\_12\catcode `\%12\relax}%
\providecommand \@@startlink[1]{}%
\providecommand \@@endlink[0]{}%
\providecommand \url  [0]{\begingroup\@sanitize@url \@url }%
\providecommand \@url [1]{\endgroup\@href {#1}{\urlprefix }}%
\providecommand \urlprefix  [0]{URL }%
\providecommand \Eprint [0]{\href }%
\providecommand \doibase [0]{http://dx.doi.org/}%
\providecommand \selectlanguage [0]{\@gobble}%
\providecommand \bibinfo  [0]{\@secondoftwo}%
\providecommand \bibfield  [0]{\@secondoftwo}%
\providecommand \translation [1]{[#1]}%
\providecommand \BibitemOpen [0]{}%
\providecommand \bibitemStop [0]{}%
\providecommand \bibitemNoStop [0]{.\EOS\space}%
\providecommand \EOS [0]{\spacefactor3000\relax}%
\providecommand \BibitemShut  [1]{\csname bibitem#1\endcsname}%
\let\auto@bib@innerbib\@empty
%</preamble>
\bibitem [{\citenamefont {Landau}(1957{\natexlab{a}})}]{Landau1957}%
  \BibitemOpen
  \bibfield  {author} {\bibinfo {author} {\bibfnamefont {LD}~\bibnamefont
  {Landau}},\ }\bibfield  {title} {\enquote {\bibinfo {title} {The theory of a
  fermi liquid},}\ }\href@noop {} {\bibfield  {journal} {\bibinfo  {journal}
  {J. Exp. Theor. Phys.}\ }\textbf {\bibinfo {volume} {3}},\ \bibinfo {pages}
  {920--925} (\bibinfo {year} {1957}{\natexlab{a}})}\BibitemShut {NoStop}%
\bibitem [{\citenamefont {Landau}(1957{\natexlab{b}})}]{Landau1957b}%
  \BibitemOpen
  \bibfield  {author} {\bibinfo {author} {\bibfnamefont {LD}~\bibnamefont
  {Landau}},\ }\bibfield  {title} {\enquote {\bibinfo {title} {Oscillations in
  a fermi liquid},}\ }\href@noop {} {\bibfield  {journal} {\bibinfo  {journal}
  {J. Exp. Theor. Phys.}\ }\textbf {\bibinfo {volume} {5}},\ \bibinfo {pages}
  {101--108} (\bibinfo {year} {1957}{\natexlab{b}})}\BibitemShut {NoStop}%
\bibitem [{\citenamefont {Baym}\ and\ \citenamefont
  {Pethick}(1991)}]{BaymPethick1991book}%
  \BibitemOpen
  \bibfield  {author} {\bibinfo {author} {\bibfnamefont {Gordon}\ \bibnamefont
  {Baym}}\ and\ \bibinfo {author} {\bibfnamefont {Christopher}\ \bibnamefont
  {Pethick}},\ }\href@noop {} {\emph {\bibinfo {title} {Landau Fermi-Liquid
  Theory: Concepts and Applications}}}\ (\bibinfo  {publisher} {Wiley-VCH},\
  \bibinfo {year} {1991})\BibitemShut {NoStop}%
\bibitem [{\citenamefont {Schrieffer}(1983)}]{Schrieffer1983}%
  \BibitemOpen
  \bibfield  {author} {\bibinfo {author} {\bibfnamefont {J.R.}\ \bibnamefont
  {Schrieffer}},\ }\href {https://books.google.dk/books?id=let7wRir74MC} {\emph
  {\bibinfo {title} {Theory Of Superconductivity}}},\ Advanced Book Program
  Series\ (\bibinfo  {publisher} {Avalon Publishing},\ \bibinfo {year}
  {1983})\BibitemShut {NoStop}%
\bibitem [{\citenamefont {Scalapino}(1995)}]{Scalapino1995}%
  \BibitemOpen
  \bibfield  {author} {\bibinfo {author} {\bibfnamefont {D.J.}\ \bibnamefont
  {Scalapino}},\ }\bibfield  {title} {\enquote {\bibinfo {title} {The case for
  dx2 − y2 pairing in the cuprate superconductors},}\ }\href {\doibase
  https://doi.org/10.1016/0370-1573(94)00086-I} {\bibfield  {journal} {\bibinfo
   {journal} {Physics Reports}\ }\textbf {\bibinfo {volume} {250}},\ \bibinfo
  {pages} {329 -- 365} (\bibinfo {year} {1995})}\BibitemShut {NoStop}%
\bibitem [{\citenamefont {Weinberg}(1995)}]{Weinberg1995}%
  \BibitemOpen
  \bibfield  {author} {\bibinfo {author} {\bibfnamefont {S.}~\bibnamefont
  {Weinberg}},\ }\href {https://books.google.dk/books?id=doeDB3\_WLvwC} {\emph
  {\bibinfo {title} {The Quantum Theory of Fields}}},\ \bibinfo {series} {The
  Quantum Theory of Fields 3 Volume Hardback Set}\ No.\ \bibinfo {number} {vb.
  1}\ (\bibinfo  {publisher} {Cambridge University Press},\ \bibinfo {year}
  {1995})\BibitemShut {NoStop}%
\bibitem [{\citenamefont {Chin}\ \emph {et~al.}(2010)\citenamefont {Chin},
  \citenamefont {Grimm}, \citenamefont {Julienne},\ and\ \citenamefont
  {Tiesinga}}]{Chin2010}%
  \BibitemOpen
  \bibfield  {author} {\bibinfo {author} {\bibfnamefont {Cheng}\ \bibnamefont
  {Chin}}, \bibinfo {author} {\bibfnamefont {Rudolf}\ \bibnamefont {Grimm}},
  \bibinfo {author} {\bibfnamefont {Paul}\ \bibnamefont {Julienne}}, \ and\
  \bibinfo {author} {\bibfnamefont {Eite}\ \bibnamefont {Tiesinga}},\
  }\bibfield  {title} {\enquote {\bibinfo {title} {Feshbach resonances in
  ultracold gases},}\ }\href {\doibase 10.1103/RevModPhys.82.1225} {\bibfield
  {journal} {\bibinfo  {journal} {Rev. Mod. Phys.}\ }\textbf {\bibinfo {volume}
  {82}},\ \bibinfo {pages} {1225--1286} (\bibinfo {year} {2010})}\BibitemShut
  {NoStop}%
\bibitem [{\citenamefont {Schirotzek}\ \emph {et~al.}(2009)\citenamefont
  {Schirotzek}, \citenamefont {Wu}, \citenamefont {Sommer},\ and\ \citenamefont
  {Zwierlein}}]{Schirotzek2009}%
  \BibitemOpen
  \bibfield  {author} {\bibinfo {author} {\bibfnamefont {Andr\'e}\ \bibnamefont
  {Schirotzek}}, \bibinfo {author} {\bibfnamefont {Cheng-Hsun}\ \bibnamefont
  {Wu}}, \bibinfo {author} {\bibfnamefont {Ariel}\ \bibnamefont {Sommer}}, \
  and\ \bibinfo {author} {\bibfnamefont {Martin~W.}\ \bibnamefont
  {Zwierlein}},\ }\bibfield  {title} {\enquote {\bibinfo {title} {Observation
  of fermi polarons in a tunable fermi liquid of ultracold atoms},}\ }\href
  {\doibase 10.1103/PhysRevLett.102.230402} {\bibfield  {journal} {\bibinfo
  {journal} {Phys. Rev. Lett.}\ }\textbf {\bibinfo {volume} {102}},\ \bibinfo
  {pages} {230402} (\bibinfo {year} {2009})}\BibitemShut {NoStop}%
\bibitem [{\citenamefont {Kohstall}\ \emph {et~al.}(2012)\citenamefont
  {Kohstall}, \citenamefont {Zaccanti}, \citenamefont {Jag}, \citenamefont
  {Trenkwalder}, \citenamefont {Massignan}, \citenamefont {Bruun},
  \citenamefont {Schreck},\ and\ \citenamefont {Grimm}}]{Kohstall2012}%
  \BibitemOpen
  \bibfield  {author} {\bibinfo {author} {\bibfnamefont {C.}~\bibnamefont
  {Kohstall}}, \bibinfo {author} {\bibfnamefont {M.}~\bibnamefont {Zaccanti}},
  \bibinfo {author} {\bibfnamefont {M.}~\bibnamefont {Jag}}, \bibinfo {author}
  {\bibfnamefont {A.}~\bibnamefont {Trenkwalder}}, \bibinfo {author}
  {\bibfnamefont {P.}~\bibnamefont {Massignan}}, \bibinfo {author}
  {\bibfnamefont {G.~M.}\ \bibnamefont {Bruun}}, \bibinfo {author}
  {\bibfnamefont {F.}~\bibnamefont {Schreck}}, \ and\ \bibinfo {author}
  {\bibfnamefont {R.}~\bibnamefont {Grimm}},\ }\bibfield  {title} {\enquote
  {\bibinfo {title} {Metastability and coherence of repulsive polarons in a
  strongly interacting fermi mixture},}\ }\href
  {http://dx.doi.org/10.1038/nature11065} {\bibfield  {journal} {\bibinfo
  {journal} {Nature}\ }\textbf {\bibinfo {volume} {485}},\ \bibinfo {pages}
  {615--618} (\bibinfo {year} {2012})}\BibitemShut {NoStop}%
\bibitem [{\citenamefont {{Koschorreck}}\ \emph {et~al.}(2012)\citenamefont
  {{Koschorreck}}, \citenamefont {{Pertot}}, \citenamefont {{Vogt}},
  \citenamefont {{Fr{\"o}hlich}}, \citenamefont {{Feld}},\ and\ \citenamefont
  {{K{\"o}hl}}}]{Koschorreck2012}%
  \BibitemOpen
  \bibfield  {author} {\bibinfo {author} {\bibfnamefont {M.}~\bibnamefont
  {{Koschorreck}}}, \bibinfo {author} {\bibfnamefont {D.}~\bibnamefont
  {{Pertot}}}, \bibinfo {author} {\bibfnamefont {E.}~\bibnamefont {{Vogt}}},
  \bibinfo {author} {\bibfnamefont {B.}~\bibnamefont {{Fr{\"o}hlich}}},
  \bibinfo {author} {\bibfnamefont {M.}~\bibnamefont {{Feld}}}, \ and\ \bibinfo
  {author} {\bibfnamefont {M.}~\bibnamefont {{K{\"o}hl}}},\ }\bibfield  {title}
  {\enquote {\bibinfo {title} {{Attractive and repulsive Fermi polarons in two
  dimensions}},}\ }\href {\doibase 10.1038/nature11151} {\bibfield  {journal}
  {\bibinfo  {journal} {Nature}\ }\textbf {\bibinfo {volume} {485}},\ \bibinfo
  {pages} {619--622} (\bibinfo {year} {2012})}\BibitemShut {NoStop}%
\bibitem [{\citenamefont {Scazza}\ \emph {et~al.}(2017)\citenamefont {Scazza},
  \citenamefont {Valtolina}, \citenamefont {Massignan}, \citenamefont {Recati},
  \citenamefont {Amico}, \citenamefont {Burchianti}, \citenamefont {Fort},
  \citenamefont {Inguscio}, \citenamefont {Zaccanti},\ and\ \citenamefont
  {Roati}}]{Scazza2017}%
  \BibitemOpen
  \bibfield  {author} {\bibinfo {author} {\bibfnamefont {F.}~\bibnamefont
  {Scazza}}, \bibinfo {author} {\bibfnamefont {G.}~\bibnamefont {Valtolina}},
  \bibinfo {author} {\bibfnamefont {P.}~\bibnamefont {Massignan}}, \bibinfo
  {author} {\bibfnamefont {A.}~\bibnamefont {Recati}}, \bibinfo {author}
  {\bibfnamefont {A.}~\bibnamefont {Amico}}, \bibinfo {author} {\bibfnamefont
  {A.}~\bibnamefont {Burchianti}}, \bibinfo {author} {\bibfnamefont
  {C.}~\bibnamefont {Fort}}, \bibinfo {author} {\bibfnamefont {M.}~\bibnamefont
  {Inguscio}}, \bibinfo {author} {\bibfnamefont {M.}~\bibnamefont {Zaccanti}},
  \ and\ \bibinfo {author} {\bibfnamefont {G.}~\bibnamefont {Roati}},\
  }\bibfield  {title} {\enquote {\bibinfo {title} {Repulsive fermi polarons in
  a resonant mixture of ultracold $^{6}\mathrm{Li}$ atoms},}\ }\href {\doibase
  10.1103/PhysRevLett.118.083602} {\bibfield  {journal} {\bibinfo  {journal}
  {Phys. Rev. Lett.}\ }\textbf {\bibinfo {volume} {118}},\ \bibinfo {pages}
  {083602} (\bibinfo {year} {2017})}\BibitemShut {NoStop}%
\bibitem [{\citenamefont {Chevy}(2006)}]{Chevy2006}%
  \BibitemOpen
  \bibfield  {author} {\bibinfo {author} {\bibfnamefont {F.}~\bibnamefont
  {Chevy}},\ }\bibfield  {title} {\enquote {\bibinfo {title} {Universal phase
  diagram of a strongly interacting fermi gas with unbalanced spin
  populations},}\ }\href {\doibase 10.1103/PhysRevA.74.063628} {\bibfield
  {journal} {\bibinfo  {journal} {Phys. Rev. A}\ }\textbf {\bibinfo {volume}
  {74}},\ \bibinfo {pages} {063628} (\bibinfo {year} {2006})}\BibitemShut
  {NoStop}%
\bibitem [{\citenamefont {Prokof'ev}\ and\ \citenamefont
  {Svistunov}(2008)}]{Prokofev2008}%
  \BibitemOpen
  \bibfield  {author} {\bibinfo {author} {\bibfnamefont {Nikolay}\ \bibnamefont
  {Prokof'ev}}\ and\ \bibinfo {author} {\bibfnamefont {Boris}\ \bibnamefont
  {Svistunov}},\ }\bibfield  {title} {\enquote {\bibinfo {title} {Fermi-polaron
  problem: Diagrammatic monte carlo method for divergent sign-alternating
  series},}\ }\href {\doibase 10.1103/PhysRevB.77.020408} {\bibfield  {journal}
  {\bibinfo  {journal} {Phys. Rev. B}\ }\textbf {\bibinfo {volume} {77}},\
  \bibinfo {pages} {020408} (\bibinfo {year} {2008})}\BibitemShut {NoStop}%
\bibitem [{\citenamefont {Mora}\ and\ \citenamefont {Chevy}(2010)}]{Mora2010}%
  \BibitemOpen
  \bibfield  {author} {\bibinfo {author} {\bibfnamefont {Christophe}\
  \bibnamefont {Mora}}\ and\ \bibinfo {author} {\bibfnamefont {Fr\'ed\'eric}\
  \bibnamefont {Chevy}},\ }\bibfield  {title} {\enquote {\bibinfo {title}
  {Normal phase of an imbalanced fermi gas},}\ }\href {\doibase
  10.1103/PhysRevLett.104.230402} {\bibfield  {journal} {\bibinfo  {journal}
  {Phys. Rev. Lett.}\ }\textbf {\bibinfo {volume} {104}},\ \bibinfo {pages}
  {230402} (\bibinfo {year} {2010})}\BibitemShut {NoStop}%
\bibitem [{\citenamefont {Punk}\ \emph {et~al.}(2009)\citenamefont {Punk},
  \citenamefont {Dumitrescu},\ and\ \citenamefont {Zwerger}}]{Punk2009}%
  \BibitemOpen
  \bibfield  {author} {\bibinfo {author} {\bibfnamefont {M.}~\bibnamefont
  {Punk}}, \bibinfo {author} {\bibfnamefont {P.~T.}\ \bibnamefont
  {Dumitrescu}}, \ and\ \bibinfo {author} {\bibfnamefont {W.}~\bibnamefont
  {Zwerger}},\ }\bibfield  {title} {\enquote {\bibinfo {title}
  {Polaron-to-molecule transition in a strongly imbalanced fermi gas},}\ }\href
  {\doibase 10.1103/PhysRevA.80.053605} {\bibfield  {journal} {\bibinfo
  {journal} {Phys. Rev. A}\ }\textbf {\bibinfo {volume} {80}},\ \bibinfo
  {pages} {053605} (\bibinfo {year} {2009})}\BibitemShut {NoStop}%
\bibitem [{\citenamefont {Combescot}\ \emph {et~al.}(2009)\citenamefont
  {Combescot}, \citenamefont {Giraud},\ and\ \citenamefont
  {Leyronas}}]{Combescot2009}%
  \BibitemOpen
  \bibfield  {author} {\bibinfo {author} {\bibfnamefont {R.}~\bibnamefont
  {Combescot}}, \bibinfo {author} {\bibfnamefont {S.}~\bibnamefont {Giraud}}, \
  and\ \bibinfo {author} {\bibfnamefont {X.}~\bibnamefont {Leyronas}},\
  }\bibfield  {title} {\enquote {\bibinfo {title} {{Analytical theory of the
  dressed bound state in highly polarized Fermi gases}},}\ }\href {\doibase
  10.1209/0295-5075/88/60007} {\bibfield  {journal} {\bibinfo  {journal} {EPL}\
  }\textbf {\bibinfo {volume} {88}},\ \bibinfo {pages} {60007} (\bibinfo {year}
  {2009})}\BibitemShut {NoStop}%
\bibitem [{\citenamefont {Cui}\ and\ \citenamefont {Zhai}(2010)}]{Cui2010}%
  \BibitemOpen
  \bibfield  {author} {\bibinfo {author} {\bibfnamefont {Xiaoling}\
  \bibnamefont {Cui}}\ and\ \bibinfo {author} {\bibfnamefont {Hui}\
  \bibnamefont {Zhai}},\ }\bibfield  {title} {\enquote {\bibinfo {title}
  {Stability of a fully magnetized ferromagnetic state in repulsively
  interacting ultracold fermi gases},}\ }\href {\doibase
  10.1103/PhysRevA.81.041602} {\bibfield  {journal} {\bibinfo  {journal} {Phys.
  Rev. A}\ }\textbf {\bibinfo {volume} {81}},\ \bibinfo {pages} {041602(R)}
  (\bibinfo {year} {2010})}\BibitemShut {NoStop}%
\bibitem [{\citenamefont {Massignan}\ and\ \citenamefont
  {Bruun}(2011)}]{Massignan2011}%
  \BibitemOpen
  \bibfield  {author} {\bibinfo {author} {\bibfnamefont {P.}~\bibnamefont
  {Massignan}}\ and\ \bibinfo {author} {\bibfnamefont {G.M.}\ \bibnamefont
  {Bruun}},\ }\bibfield  {title} {\enquote {\bibinfo {title} {Repulsive
  polarons and itinerant ferromagnetism in strongly polarized fermi gases},}\
  }\href {\doibase 10.1140/epjd/e2011-20084-5} {\bibfield  {journal} {\bibinfo
  {journal} {EPJ D}\ }\textbf {\bibinfo {volume} {65}},\ \bibinfo {pages}
  {83--89} (\bibinfo {year} {2011})}\BibitemShut {NoStop}%
\bibitem [{\citenamefont {Massignan}\ \emph {et~al.}(2014)\citenamefont
  {Massignan}, \citenamefont {Zaccanti},\ and\ \citenamefont
  {Bruun}}]{Massignan2014}%
  \BibitemOpen
  \bibfield  {author} {\bibinfo {author} {\bibfnamefont {Pietro}\ \bibnamefont
  {Massignan}}, \bibinfo {author} {\bibfnamefont {Matteo}\ \bibnamefont
  {Zaccanti}}, \ and\ \bibinfo {author} {\bibfnamefont {Georg~M}\ \bibnamefont
  {Bruun}},\ }\bibfield  {title} {\enquote {\bibinfo {title} {Polarons, dressed
  molecules and itinerant ferromagnetism in ultracold fermi gases},}\ }\href
  {http://stacks.iop.org/0034-4885/77/i=3/a=034401} {\bibfield  {journal}
  {\bibinfo  {journal} {Rep. Progr. Phys.}\ }\textbf {\bibinfo {volume} {77}},\
  \bibinfo {pages} {034401} (\bibinfo {year} {2014})}\BibitemShut {NoStop}%
\bibitem [{\citenamefont {J\o{}rgensen}\ \emph {et~al.}(2016)\citenamefont
  {J\o{}rgensen}, \citenamefont {Wacker}, \citenamefont {Skalmstang},
  \citenamefont {Parish}, \citenamefont {Levinsen}, \citenamefont
  {Christensen}, \citenamefont {Bruun},\ and\ \citenamefont
  {Arlt}}]{Jorgensen2016}%
  \BibitemOpen
  \bibfield  {author} {\bibinfo {author} {\bibfnamefont {Nils~B.}\ \bibnamefont
  {J\o{}rgensen}}, \bibinfo {author} {\bibfnamefont {Lars}\ \bibnamefont
  {Wacker}}, \bibinfo {author} {\bibfnamefont {Kristoffer~T.}\ \bibnamefont
  {Skalmstang}}, \bibinfo {author} {\bibfnamefont {Meera~M.}\ \bibnamefont
  {Parish}}, \bibinfo {author} {\bibfnamefont {Jesper}\ \bibnamefont
  {Levinsen}}, \bibinfo {author} {\bibfnamefont {Rasmus~S.}\ \bibnamefont
  {Christensen}}, \bibinfo {author} {\bibfnamefont {Georg~M.}\ \bibnamefont
  {Bruun}}, \ and\ \bibinfo {author} {\bibfnamefont {Jan~J.}\ \bibnamefont
  {Arlt}},\ }\bibfield  {title} {\enquote {\bibinfo {title} {Observation of
  attractive and repulsive polarons in a bose-einstein condensate},}\ }\href
  {\doibase 10.1103/PhysRevLett.117.055302} {\bibfield  {journal} {\bibinfo
  {journal} {Phys. Rev. Lett.}\ }\textbf {\bibinfo {volume} {117}},\ \bibinfo
  {pages} {055302} (\bibinfo {year} {2016})}\BibitemShut {NoStop}%
\bibitem [{\citenamefont {Hu}\ \emph {et~al.}(2016)\citenamefont {Hu},
  \citenamefont {Van~de Graaff}, \citenamefont {Kedar}, \citenamefont {Corson},
  \citenamefont {Cornell},\ and\ \citenamefont {Jin}}]{Hu2016}%
  \BibitemOpen
  \bibfield  {author} {\bibinfo {author} {\bibfnamefont {Ming-Guang}\
  \bibnamefont {Hu}}, \bibinfo {author} {\bibfnamefont {Michael~J.}\
  \bibnamefont {Van~de Graaff}}, \bibinfo {author} {\bibfnamefont {Dhruv}\
  \bibnamefont {Kedar}}, \bibinfo {author} {\bibfnamefont {John~P.}\
  \bibnamefont {Corson}}, \bibinfo {author} {\bibfnamefont {Eric~A.}\
  \bibnamefont {Cornell}}, \ and\ \bibinfo {author} {\bibfnamefont
  {Deborah~S.}\ \bibnamefont {Jin}},\ }\bibfield  {title} {\enquote {\bibinfo
  {title} {Bose polarons in the strongly interacting regime},}\ }\href
  {\doibase 10.1103/PhysRevLett.117.055301} {\bibfield  {journal} {\bibinfo
  {journal} {Phys. Rev. Lett.}\ }\textbf {\bibinfo {volume} {117}},\ \bibinfo
  {pages} {055301} (\bibinfo {year} {2016})}\BibitemShut {NoStop}%
\bibitem [{\citenamefont {Catani}\ \emph {et~al.}(2012)\citenamefont {Catani},
  \citenamefont {Lamporesi}, \citenamefont {Naik}, \citenamefont {Gring},
  \citenamefont {Inguscio}, \citenamefont {Minardi}, \citenamefont {Kantian},\
  and\ \citenamefont {Giamarchi}}]{Catani2012}%
  \BibitemOpen
  \bibfield  {author} {\bibinfo {author} {\bibfnamefont {J.}~\bibnamefont
  {Catani}}, \bibinfo {author} {\bibfnamefont {G.}~\bibnamefont {Lamporesi}},
  \bibinfo {author} {\bibfnamefont {D.}~\bibnamefont {Naik}}, \bibinfo {author}
  {\bibfnamefont {M.}~\bibnamefont {Gring}}, \bibinfo {author} {\bibfnamefont
  {M.}~\bibnamefont {Inguscio}}, \bibinfo {author} {\bibfnamefont
  {F.}~\bibnamefont {Minardi}}, \bibinfo {author} {\bibfnamefont
  {A.}~\bibnamefont {Kantian}}, \ and\ \bibinfo {author} {\bibfnamefont
  {T.}~\bibnamefont {Giamarchi}},\ }\bibfield  {title} {\enquote {\bibinfo
  {title} {Quantum dynamics of impurities in a one-dimensional bose gas},}\
  }\href {\doibase 10.1103/PhysRevA.85.023623} {\bibfield  {journal} {\bibinfo
  {journal} {Phys. Rev. A}\ }\textbf {\bibinfo {volume} {85}},\ \bibinfo
  {pages} {023623} (\bibinfo {year} {2012})}\BibitemShut {NoStop}%
\bibitem [{\citenamefont {Li}\ and\ \citenamefont {Das~Sarma}(2014)}]{Li2014}%
  \BibitemOpen
  \bibfield  {author} {\bibinfo {author} {\bibfnamefont {Weiran}\ \bibnamefont
  {Li}}\ and\ \bibinfo {author} {\bibfnamefont {S.}~\bibnamefont {Das~Sarma}},\
  }\bibfield  {title} {\enquote {\bibinfo {title} {Variational study of
  polarons in bose-einstein condensates},}\ }\href {\doibase
  10.1103/PhysRevA.90.013618} {\bibfield  {journal} {\bibinfo  {journal} {Phys.
  Rev. A}\ }\textbf {\bibinfo {volume} {90}},\ \bibinfo {pages} {013618}
  (\bibinfo {year} {2014})}\BibitemShut {NoStop}%
\bibitem [{\citenamefont {Levinsen}\ \emph {et~al.}(2015)\citenamefont
  {Levinsen}, \citenamefont {Parish},\ and\ \citenamefont
  {Bruun}}]{Levinsen2015}%
  \BibitemOpen
  \bibfield  {author} {\bibinfo {author} {\bibfnamefont {Jesper}\ \bibnamefont
  {Levinsen}}, \bibinfo {author} {\bibfnamefont {Meera~M.}\ \bibnamefont
  {Parish}}, \ and\ \bibinfo {author} {\bibfnamefont {Georg~M.}\ \bibnamefont
  {Bruun}},\ }\bibfield  {title} {\enquote {\bibinfo {title} {Impurity in a
  bose-einstein condensate and the efimov effect},}\ }\href {\doibase
  10.1103/PhysRevLett.115.125302} {\bibfield  {journal} {\bibinfo  {journal}
  {Phys. Rev. Lett.}\ }\textbf {\bibinfo {volume} {115}},\ \bibinfo {pages}
  {125302} (\bibinfo {year} {2015})}\BibitemShut {NoStop}%
\bibitem [{\citenamefont {Shchadilova}\ \emph {et~al.}(2016)\citenamefont
  {Shchadilova}, \citenamefont {Schmidt}, \citenamefont {Grusdt},\ and\
  \citenamefont {Demler}}]{Shchadilova2016}%
  \BibitemOpen
  \bibfield  {author} {\bibinfo {author} {\bibfnamefont {Yulia~E.}\
  \bibnamefont {Shchadilova}}, \bibinfo {author} {\bibfnamefont {Richard}\
  \bibnamefont {Schmidt}}, \bibinfo {author} {\bibfnamefont {Fabian}\
  \bibnamefont {Grusdt}}, \ and\ \bibinfo {author} {\bibfnamefont {Eugene}\
  \bibnamefont {Demler}},\ }\bibfield  {title} {\enquote {\bibinfo {title}
  {Quantum dynamics of ultracold bose polarons},}\ }\href {\doibase
  10.1103/PhysRevLett.117.113002} {\bibfield  {journal} {\bibinfo  {journal}
  {Phys. Rev. Lett.}\ }\textbf {\bibinfo {volume} {117}},\ \bibinfo {pages}
  {113002} (\bibinfo {year} {2016})}\BibitemShut {NoStop}%
\bibitem [{\citenamefont {Tempere}\ \emph {et~al.}(2009)\citenamefont
  {Tempere}, \citenamefont {Casteels}, \citenamefont {Oberthaler},
  \citenamefont {Knoop}, \citenamefont {Timmermans},\ and\ \citenamefont
  {Devreese}}]{Tempere2009}%
  \BibitemOpen
  \bibfield  {author} {\bibinfo {author} {\bibfnamefont {J.}~\bibnamefont
  {Tempere}}, \bibinfo {author} {\bibfnamefont {W.}~\bibnamefont {Casteels}},
  \bibinfo {author} {\bibfnamefont {M.~K.}\ \bibnamefont {Oberthaler}},
  \bibinfo {author} {\bibfnamefont {S.}~\bibnamefont {Knoop}}, \bibinfo
  {author} {\bibfnamefont {E.}~\bibnamefont {Timmermans}}, \ and\ \bibinfo
  {author} {\bibfnamefont {J.~T.}\ \bibnamefont {Devreese}},\ }\bibfield
  {title} {\enquote {\bibinfo {title} {Feynman path-integral treatment of the
  bec-impurity polaron},}\ }\href {\doibase 10.1103/PhysRevB.80.184504}
  {\bibfield  {journal} {\bibinfo  {journal} {Phys. Rev. B}\ }\textbf {\bibinfo
  {volume} {80}},\ \bibinfo {pages} {184504} (\bibinfo {year}
  {2009})}\BibitemShut {NoStop}%
\bibitem [{\citenamefont {Rath}\ and\ \citenamefont
  {Schmidt}(2013)}]{Rath2013}%
  \BibitemOpen
  \bibfield  {author} {\bibinfo {author} {\bibfnamefont {Steffen~Patrick}\
  \bibnamefont {Rath}}\ and\ \bibinfo {author} {\bibfnamefont {Richard}\
  \bibnamefont {Schmidt}},\ }\bibfield  {title} {\enquote {\bibinfo {title}
  {Field-theoretical study of the bose polaron},}\ }\href {\doibase
  10.1103/PhysRevA.88.053632} {\bibfield  {journal} {\bibinfo  {journal} {Phys.
  Rev. A}\ }\textbf {\bibinfo {volume} {88}},\ \bibinfo {pages} {053632}
  (\bibinfo {year} {2013})}\BibitemShut {NoStop}%
\bibitem [{\citenamefont {Christensen}\ \emph {et~al.}(2015)\citenamefont
  {Christensen}, \citenamefont {Levinsen},\ and\ \citenamefont
  {Bruun}}]{Christensen2015}%
  \BibitemOpen
  \bibfield  {author} {\bibinfo {author} {\bibfnamefont {Rasmus~S\o{}gaard}\
  \bibnamefont {Christensen}}, \bibinfo {author} {\bibfnamefont {Jesper}\
  \bibnamefont {Levinsen}}, \ and\ \bibinfo {author} {\bibfnamefont {Georg~M.}\
  \bibnamefont {Bruun}},\ }\bibfield  {title} {\enquote {\bibinfo {title}
  {Quasiparticle properties of a mobile impurity in a bose-einstein
  condensate},}\ }\href {\doibase 10.1103/PhysRevLett.115.160401} {\bibfield
  {journal} {\bibinfo  {journal} {Phys. Rev. Lett.}\ }\textbf {\bibinfo
  {volume} {115}},\ \bibinfo {pages} {160401} (\bibinfo {year}
  {2015})}\BibitemShut {NoStop}%
\bibitem [{\citenamefont {Grusdt}\ \emph {et~al.}(2017)\citenamefont {Grusdt},
  \citenamefont {Schmidt}, \citenamefont {Shchadilova},\ and\ \citenamefont
  {Demler}}]{Grusdt2017}%
  \BibitemOpen
  \bibfield  {author} {\bibinfo {author} {\bibfnamefont {F.}~\bibnamefont
  {Grusdt}}, \bibinfo {author} {\bibfnamefont {R.}~\bibnamefont {Schmidt}},
  \bibinfo {author} {\bibfnamefont {Y.~E.}\ \bibnamefont {Shchadilova}}, \ and\
  \bibinfo {author} {\bibfnamefont {E.}~\bibnamefont {Demler}},\ }\bibfield
  {title} {\enquote {\bibinfo {title} {Strong-coupling bose polarons in a
  bose-einstein condensate},}\ }\href {\doibase 10.1103/PhysRevA.96.013607}
  {\bibfield  {journal} {\bibinfo  {journal} {Phys. Rev. A}\ }\textbf {\bibinfo
  {volume} {96}},\ \bibinfo {pages} {013607} (\bibinfo {year}
  {2017})}\BibitemShut {NoStop}%
\bibitem [{\citenamefont {Astrakharchik}\ and\ \citenamefont
  {Pitaevskii}(2004)}]{Astrakharchik2004}%
  \BibitemOpen
  \bibfield  {author} {\bibinfo {author} {\bibfnamefont {G.~E.}\ \bibnamefont
  {Astrakharchik}}\ and\ \bibinfo {author} {\bibfnamefont {L.~P.}\ \bibnamefont
  {Pitaevskii}},\ }\bibfield  {title} {\enquote {\bibinfo {title} {Motion of a
  heavy impurity through a bose-einstein condensate},}\ }\href {\doibase
  10.1103/PhysRevA.70.013608} {\bibfield  {journal} {\bibinfo  {journal} {Phys.
  Rev. A}\ }\textbf {\bibinfo {volume} {70}},\ \bibinfo {pages} {013608}
  (\bibinfo {year} {2004})}\BibitemShut {NoStop}%
\bibitem [{\citenamefont {Cucchietti}\ and\ \citenamefont
  {Timmermans}(2006)}]{Cucchietti2006}%
  \BibitemOpen
  \bibfield  {author} {\bibinfo {author} {\bibfnamefont {F.~M.}\ \bibnamefont
  {Cucchietti}}\ and\ \bibinfo {author} {\bibfnamefont {E.}~\bibnamefont
  {Timmermans}},\ }\bibfield  {title} {\enquote {\bibinfo {title}
  {Strong-coupling polarons in dilute gas bose-einstein condensates},}\ }\href
  {\doibase 10.1103/PhysRevLett.96.210401} {\bibfield  {journal} {\bibinfo
  {journal} {Phys. Rev. Lett.}\ }\textbf {\bibinfo {volume} {96}},\ \bibinfo
  {pages} {210401} (\bibinfo {year} {2006})}\BibitemShut {NoStop}%
\bibitem [{\citenamefont {Volosniev}\ \emph {et~al.}(2015)\citenamefont
  {Volosniev}, \citenamefont {Hammer},\ and\ \citenamefont
  {Zinner}}]{Volosniev2015}%
  \BibitemOpen
  \bibfield  {author} {\bibinfo {author} {\bibfnamefont {A.~G.}\ \bibnamefont
  {Volosniev}}, \bibinfo {author} {\bibfnamefont {H.-W.}\ \bibnamefont
  {Hammer}}, \ and\ \bibinfo {author} {\bibfnamefont {N.~T.}\ \bibnamefont
  {Zinner}},\ }\bibfield  {title} {\enquote {\bibinfo {title} {Real-time
  dynamics of an impurity in an ideal bose gas in a trap},}\ }\href {\doibase
  10.1103/PhysRevA.92.023623} {\bibfield  {journal} {\bibinfo  {journal} {Phys.
  Rev. A}\ }\textbf {\bibinfo {volume} {92}},\ \bibinfo {pages} {023623}
  (\bibinfo {year} {2015})}\BibitemShut {NoStop}%
\bibitem [{\citenamefont {Schmidt}\ \emph {et~al.}(2016)\citenamefont
  {Schmidt}, \citenamefont {Sadeghpour},\ and\ \citenamefont
  {Demler}}]{Schmidt2016}%
  \BibitemOpen
  \bibfield  {author} {\bibinfo {author} {\bibfnamefont {Richard}\ \bibnamefont
  {Schmidt}}, \bibinfo {author} {\bibfnamefont {H.~R.}\ \bibnamefont
  {Sadeghpour}}, \ and\ \bibinfo {author} {\bibfnamefont {E.}~\bibnamefont
  {Demler}},\ }\bibfield  {title} {\enquote {\bibinfo {title} {Mesoscopic
  rydberg impurity in an atomic quantum gas},}\ }\href {\doibase
  10.1103/PhysRevLett.116.105302} {\bibfield  {journal} {\bibinfo  {journal}
  {Phys. Rev. Lett.}\ }\textbf {\bibinfo {volume} {116}},\ \bibinfo {pages}
  {105302} (\bibinfo {year} {2016})}\BibitemShut {NoStop}%
\bibitem [{\citenamefont {Pe{\~n}a~Ardila}\ and\ \citenamefont
  {Giorgini}(2015)}]{Ardila2015}%
  \BibitemOpen
  \bibfield  {author} {\bibinfo {author} {\bibfnamefont {L.~A.}\ \bibnamefont
  {Pe{\~n}a~Ardila}}\ and\ \bibinfo {author} {\bibfnamefont {S.}~\bibnamefont
  {Giorgini}},\ }\bibfield  {title} {\enquote {\bibinfo {title} {{Impurity in a
  Bose-Einstein condensate: Study of the attractive and repulsive branch using
  quantum Monte Carlo methods}},}\ }\href
  {http://dx.doi.org/10.1103/PhysRevA.92.033612} {\bibfield  {journal}
  {\bibinfo  {journal} {Phys. Rev. A}\ }\textbf {\bibinfo {volume} {92}},\
  \bibinfo {pages} {033612} (\bibinfo {year} {2015})}\BibitemShut {NoStop}%
\bibitem [{\citenamefont {Pe{\~n}a~Ardila}\ and\ \citenamefont
  {Giorgini}(2016)}]{Ardila2016}%
  \BibitemOpen
  \bibfield  {author} {\bibinfo {author} {\bibfnamefont {L.~A.}\ \bibnamefont
  {Pe{\~n}a~Ardila}}\ and\ \bibinfo {author} {\bibfnamefont {S.}~\bibnamefont
  {Giorgini}},\ }\bibfield  {title} {\enquote {\bibinfo {title} {Bose polaron
  problem: Effect of mass imbalance on binding energy},}\ }\href {\doibase
  10.1103/PhysRevA.94.063640} {\bibfield  {journal} {\bibinfo  {journal} {Phys.
  Rev. A}\ }\textbf {\bibinfo {volume} {94}},\ \bibinfo {pages} {063640}
  (\bibinfo {year} {2016})}\BibitemShut {NoStop}%
\bibitem [{\citenamefont {Yu}\ and\ \citenamefont {Pethick}(2012)}]{Yu2012a}%
  \BibitemOpen
  \bibfield  {author} {\bibinfo {author} {\bibfnamefont {Zhenhua}\ \bibnamefont
  {Yu}}\ and\ \bibinfo {author} {\bibfnamefont {C.~J.}\ \bibnamefont
  {Pethick}},\ }\bibfield  {title} {\enquote {\bibinfo {title} {Induced
  interactions in dilute atomic gases and liquid helium mixtures},}\ }\href
  {\doibase 10.1103/PhysRevA.85.063616} {\bibfield  {journal} {\bibinfo
  {journal} {Phys. Rev. A}\ }\textbf {\bibinfo {volume} {85}},\ \bibinfo
  {pages} {063616} (\bibinfo {year} {2012})}\BibitemShut {NoStop}%
\bibitem [{\citenamefont {{Hu}}\ \emph {et~al.}(2017)\citenamefont {{Hu}},
  \citenamefont {{Mulkerin}}, \citenamefont {{Wang}},\ and\ \citenamefont
  {{Liu}}}]{HuM2017}%
  \BibitemOpen
  \bibfield  {author} {\bibinfo {author} {\bibfnamefont {H.}~\bibnamefont
  {{Hu}}}, \bibinfo {author} {\bibfnamefont {B.~C.}\ \bibnamefont
  {{Mulkerin}}}, \bibinfo {author} {\bibfnamefont {J.}~\bibnamefont {{Wang}}},
  \ and\ \bibinfo {author} {\bibfnamefont {X.-J.}\ \bibnamefont {{Liu}}},\
  }\bibfield  {title} {\enquote {\bibinfo {title} {{Attractive Fermi polarons
  at nonzero temperature with finite impurityconcentration}},}\ }\href@noop {}
  {\bibfield  {journal} {\bibinfo  {journal} {ArXiv e-prints}\ } (\bibinfo
  {year} {2017})},\ \Eprint {http://arxiv.org/abs/1708.03410} {arXiv:1708.03410
  [cond-mat.quant-gas]} \BibitemShut {NoStop}%
\bibitem [{\citenamefont {{Dehkharghani}}\ \emph {et~al.}(2017)\citenamefont
  {{Dehkharghani}}, \citenamefont {{Volosniev}},\ and\ \citenamefont
  {{Zinner}}}]{Dehkharghani2017}%
  \BibitemOpen
  \bibfield  {author} {\bibinfo {author} {\bibfnamefont {A.~S.}\ \bibnamefont
  {{Dehkharghani}}}, \bibinfo {author} {\bibfnamefont {A.~G.}\ \bibnamefont
  {{Volosniev}}}, \ and\ \bibinfo {author} {\bibfnamefont {N.~T.}\ \bibnamefont
  {{Zinner}}},\ }\bibfield  {title} {\enquote {\bibinfo {title}
  {{Interaction-driven Coalescence of Two Impurities in a One-dimensional Bose
  Gas}},}\ }\href@noop {} {\bibfield  {journal} {\bibinfo  {journal} {ArXiv
  e-prints}\ } (\bibinfo {year} {2017})},\ \Eprint
  {http://arxiv.org/abs/1712.01538} {arXiv:1712.01538 [cond-mat.quant-gas]}
  \BibitemShut {NoStop}%
\bibitem [{\citenamefont {Klein}\ and\ \citenamefont
  {Fleischhauer}(2005)}]{Klein2005}%
  \BibitemOpen
  \bibfield  {author} {\bibinfo {author} {\bibfnamefont {Alexander}\
  \bibnamefont {Klein}}\ and\ \bibinfo {author} {\bibfnamefont {Michael}\
  \bibnamefont {Fleischhauer}},\ }\bibfield  {title} {\enquote {\bibinfo
  {title} {Interaction of impurity atoms in bose-einstein condensates},}\
  }\href {\doibase 10.1103/PhysRevA.71.033605} {\bibfield  {journal} {\bibinfo
  {journal} {Phys. Rev. A}\ }\textbf {\bibinfo {volume} {71}},\ \bibinfo
  {pages} {033605} (\bibinfo {year} {2005})}\BibitemShut {NoStop}%
\bibitem [{\citenamefont {{Naidon}}(2016)}]{Naidon2016}%
  \BibitemOpen
  \bibfield  {author} {\bibinfo {author} {\bibfnamefont {P.}~\bibnamefont
  {{Naidon}}},\ }\bibfield  {title} {\enquote {\bibinfo {title} {{Two
  impurities in a Bose-Einstein condensate: from Yukawa to Efimov attracted
  polarons}},}\ }\href@noop {} {\bibfield  {journal} {\bibinfo  {journal}
  {ArXiv e-prints}\ } (\bibinfo {year} {2016})},\ \Eprint
  {http://arxiv.org/abs/1607.04507} {arXiv:1607.04507 [cond-mat.quant-gas]}
  \BibitemShut {NoStop}%
\bibitem [{\citenamefont {Levinsen}\ \emph {et~al.}(2017)\citenamefont
  {Levinsen}, \citenamefont {Parish}, \citenamefont {Christensen},
  \citenamefont {Arlt},\ and\ \citenamefont {Bruun}}]{Levinsen2017}%
  \BibitemOpen
  \bibfield  {author} {\bibinfo {author} {\bibfnamefont {Jesper}\ \bibnamefont
  {Levinsen}}, \bibinfo {author} {\bibfnamefont {Meera~M.}\ \bibnamefont
  {Parish}}, \bibinfo {author} {\bibfnamefont {Rasmus~S.}\ \bibnamefont
  {Christensen}}, \bibinfo {author} {\bibfnamefont {Jan~J.}\ \bibnamefont
  {Arlt}}, \ and\ \bibinfo {author} {\bibfnamefont {Georg~M.}\ \bibnamefont
  {Bruun}},\ }\bibfield  {title} {\enquote {\bibinfo {title}
  {Finite-temperature behavior of the bose polaron},}\ }\href {\doibase
  10.1103/PhysRevA.96.063622} {\bibfield  {journal} {\bibinfo  {journal} {Phys.
  Rev. A}\ }\textbf {\bibinfo {volume} {96}},\ \bibinfo {pages} {063622}
  (\bibinfo {year} {2017})}\BibitemShut {NoStop}%
\bibitem [{\citenamefont {Yu}\ \emph {et~al.}(2010)\citenamefont {Yu},
  \citenamefont {Z\"ollner},\ and\ \citenamefont {Pethick}}]{Yu2010}%
  \BibitemOpen
  \bibfield  {author} {\bibinfo {author} {\bibfnamefont {Zhenhua}\ \bibnamefont
  {Yu}}, \bibinfo {author} {\bibfnamefont {Sascha}\ \bibnamefont {Z\"ollner}},
  \ and\ \bibinfo {author} {\bibfnamefont {C.~J.}\ \bibnamefont {Pethick}},\
  }\bibfield  {title} {\enquote {\bibinfo {title} {Comment on ``normal phase of
  an imbalanced fermi gas''},}\ }\href {\doibase
  10.1103/PhysRevLett.105.188901} {\bibfield  {journal} {\bibinfo  {journal}
  {Phys. Rev. Lett.}\ }\textbf {\bibinfo {volume} {105}},\ \bibinfo {pages}
  {188901} (\bibinfo {year} {2010})}\BibitemShut {NoStop}%
\bibitem [{foo()}]{footnote}%
  \BibitemOpen
  \href@noop {} {}\bibinfo {note} {Technically, this term comes from the pole
  of the induced interaction, which does not occur in an exchange diagram with
  an ordinary frequency independent interaction}\BibitemShut {NoStop}%
\bibitem [{\citenamefont {Viverit}\ \emph {et~al.}(2000)\citenamefont
  {Viverit}, \citenamefont {Pethick},\ and\ \citenamefont
  {Smith}}]{Viverit2000}%
  \BibitemOpen
  \bibfield  {author} {\bibinfo {author} {\bibfnamefont {L.}~\bibnamefont
  {Viverit}}, \bibinfo {author} {\bibfnamefont {C.~J.}\ \bibnamefont
  {Pethick}}, \ and\ \bibinfo {author} {\bibfnamefont {H.}~\bibnamefont
  {Smith}},\ }\bibfield  {title} {\enquote {\bibinfo {title} {Zero-temperature
  phase diagram of binary boson-fermion mixtures},}\ }\href {\doibase
  10.1103/PhysRevA.61.053605} {\bibfield  {journal} {\bibinfo  {journal} {Phys.
  Rev. A}\ }\textbf {\bibinfo {volume} {61}},\ \bibinfo {pages} {053605}
  (\bibinfo {year} {2000})}\BibitemShut {NoStop}%
\bibitem [{\citenamefont {Shi}\ and\ \citenamefont {Griffin}(1998)}]{Shi1998}%
  \BibitemOpen
  \bibfield  {author} {\bibinfo {author} {\bibfnamefont {Hua}\ \bibnamefont
  {Shi}}\ and\ \bibinfo {author} {\bibfnamefont {Allan}\ \bibnamefont
  {Griffin}},\ }\bibfield  {title} {\enquote {\bibinfo {title}
  {Finite-temperature excitations in a dilute bose-condensed gas},}\ }\href
  {\doibase http://dx.doi.org/10.1016/S0370-1573(98)00015-5} {\bibfield
  {journal} {\bibinfo  {journal} {Phys. Rep.}\ }\textbf {\bibinfo {volume}
  {304}},\ \bibinfo {pages} {1 -- 87} (\bibinfo {year} {1998})}\BibitemShut
  {NoStop}%
\bibitem [{\citenamefont {Watabe}\ and\ \citenamefont
  {Ohashi}(2013)}]{Watabe2013}%
  \BibitemOpen
  \bibfield  {author} {\bibinfo {author} {\bibfnamefont {Shohei}\ \bibnamefont
  {Watabe}}\ and\ \bibinfo {author} {\bibfnamefont {Yoji}\ \bibnamefont
  {Ohashi}},\ }\bibfield  {title} {\enquote {\bibinfo {title} {Comparative
  studies of many-body corrections to an interacting bose-einstein
  condensate},}\ }\href {\doibase 10.1103/PhysRevA.88.053633} {\bibfield
  {journal} {\bibinfo  {journal} {Phys. Rev. A}\ }\textbf {\bibinfo {volume}
  {88}},\ \bibinfo {pages} {053633} (\bibinfo {year} {2013})}\BibitemShut
  {NoStop}%
\bibitem [{\citenamefont {Fritsch}\ \emph {et~al.}(2018)\citenamefont
  {Fritsch}, \citenamefont {Tavares}, \citenamefont {Vivanco}, \citenamefont
  {Telles}, \citenamefont {Bagnato},\ and\ \citenamefont {Henn}}]{Fritsch2018}%
  \BibitemOpen
  \bibfield  {author} {\bibinfo {author} {\bibfnamefont {A~R}\ \bibnamefont
  {Fritsch}}, \bibinfo {author} {\bibfnamefont {P~E~S}\ \bibnamefont
  {Tavares}}, \bibinfo {author} {\bibfnamefont {F~A~J}\ \bibnamefont
  {Vivanco}}, \bibinfo {author} {\bibfnamefont {G~D}\ \bibnamefont {Telles}},
  \bibinfo {author} {\bibfnamefont {V~S}\ \bibnamefont {Bagnato}}, \ and\
  \bibinfo {author} {\bibfnamefont {E~A~L}\ \bibnamefont {Henn}},\ }\bibfield
  {title} {\enquote {\bibinfo {title} {Thermodynamic measurement of the sound
  velocity of a bose gas across the transition to bose–einstein
  condensation},}\ }\href {http://stacks.iop.org/1742-5468/2018/i=5/a=053108}
  {\bibfield  {journal} {\bibinfo  {journal} {Journal of Statistical Mechanics:
  Theory and Experiment}\ }\textbf {\bibinfo {volume} {2018}},\ \bibinfo
  {pages} {053108} (\bibinfo {year} {2018})}\BibitemShut {NoStop}%
\bibitem [{\citenamefont {Lous}\ \emph {et~al.}(2017)\citenamefont {Lous},
  \citenamefont {Fritsche}, \citenamefont {Jag}, \citenamefont {Huang},\ and\
  \citenamefont {Grimm}}]{Rianne2017}%
  \BibitemOpen
  \bibfield  {author} {\bibinfo {author} {\bibfnamefont {Rianne~S.}\
  \bibnamefont {Lous}}, \bibinfo {author} {\bibfnamefont {Isabella}\
  \bibnamefont {Fritsche}}, \bibinfo {author} {\bibfnamefont {Michael}\
  \bibnamefont {Jag}}, \bibinfo {author} {\bibfnamefont {Bo}~\bibnamefont
  {Huang}}, \ and\ \bibinfo {author} {\bibfnamefont {Rudolf}\ \bibnamefont
  {Grimm}},\ }\bibfield  {title} {\enquote {\bibinfo {title} {Thermometry of a
  deeply degenerate fermi gas with a bose-einstein condensate},}\ }\href
  {\doibase 10.1103/PhysRevA.95.053627} {\bibfield  {journal} {\bibinfo
  {journal} {Phys. Rev. A}\ }\textbf {\bibinfo {volume} {95}},\ \bibinfo
  {pages} {053627} (\bibinfo {year} {2017})}\BibitemShut {NoStop}%
\bibitem [{\citenamefont {Guidini}\ \emph {et~al.}(2015)\citenamefont
  {Guidini}, \citenamefont {Bertaina}, \citenamefont {Galli},\ and\
  \citenamefont {Pieri}}]{Guidini2015}%
  \BibitemOpen
  \bibfield  {author} {\bibinfo {author} {\bibfnamefont {Andrea}\ \bibnamefont
  {Guidini}}, \bibinfo {author} {\bibfnamefont {Gianluca}\ \bibnamefont
  {Bertaina}}, \bibinfo {author} {\bibfnamefont {Davide~Emilio}\ \bibnamefont
  {Galli}}, \ and\ \bibinfo {author} {\bibfnamefont {Pierbiagio}\ \bibnamefont
  {Pieri}},\ }\bibfield  {title} {\enquote {\bibinfo {title} {Condensed phase
  of bose-fermi mixtures with a pairing interaction},}\ }\href {\doibase
  10.1103/PhysRevA.91.023603} {\bibfield  {journal} {\bibinfo  {journal} {Phys.
  Rev. A}\ }\textbf {\bibinfo {volume} {91}},\ \bibinfo {pages} {023603}
  (\bibinfo {year} {2015})}\BibitemShut {NoStop}%
\bibitem [{\citenamefont {Guenther}\ \emph {et~al.}(2018)\citenamefont
  {Guenther}, \citenamefont {Massignan}, \citenamefont {Lewenstein},\ and\
  \citenamefont {Bruun}}]{Guenther2018}%
  \BibitemOpen
  \bibfield  {author} {\bibinfo {author} {\bibfnamefont {Nils-Eric}\
  \bibnamefont {Guenther}}, \bibinfo {author} {\bibfnamefont {Pietro}\
  \bibnamefont {Massignan}}, \bibinfo {author} {\bibfnamefont {Maciej}\
  \bibnamefont {Lewenstein}}, \ and\ \bibinfo {author} {\bibfnamefont
  {Georg~M.}\ \bibnamefont {Bruun}},\ }\bibfield  {title} {\enquote {\bibinfo
  {title} {Bose polarons at finite temperature and strong coupling},}\ }\href
  {\doibase 10.1103/PhysRevLett.120.050405} {\bibfield  {journal} {\bibinfo
  {journal} {Phys. Rev. Lett.}\ }\textbf {\bibinfo {volume} {120}},\ \bibinfo
  {pages} {050405} (\bibinfo {year} {2018})}\BibitemShut {NoStop}%
\bibitem [{\citenamefont {Suchet}\ \emph {et~al.}(2017)\citenamefont {Suchet},
  \citenamefont {Wu}, \citenamefont {Chevy},\ and\ \citenamefont
  {Bruun}}]{Suchet2017}%
  \BibitemOpen
  \bibfield  {author} {\bibinfo {author} {\bibfnamefont {Daniel}\ \bibnamefont
  {Suchet}}, \bibinfo {author} {\bibfnamefont {Zhigang}\ \bibnamefont {Wu}},
  \bibinfo {author} {\bibfnamefont {Fr\'ed\'eric}\ \bibnamefont {Chevy}}, \
  and\ \bibinfo {author} {\bibfnamefont {Georg~M.}\ \bibnamefont {Bruun}},\
  }\bibfield  {title} {\enquote {\bibinfo {title} {Long-range mediated
  interactions in a mixed-dimensional system},}\ }\href {\doibase
  10.1103/PhysRevA.95.043643} {\bibfield  {journal} {\bibinfo  {journal} {Phys.
  Rev. A}\ }\textbf {\bibinfo {volume} {95}},\ \bibinfo {pages} {043643}
  (\bibinfo {year} {2017})}\BibitemShut {NoStop}%
\bibitem [{\citenamefont {Wu}\ \emph {et~al.}(2017)\citenamefont {Wu},
  \citenamefont {Yao}, \citenamefont {Chen}, \citenamefont {Liu}, \citenamefont
  {Wang}, \citenamefont {Chen},\ and\ \citenamefont {Pan}}]{YuPing2017}%
  \BibitemOpen
  \bibfield  {author} {\bibinfo {author} {\bibfnamefont {Yu-Ping}\ \bibnamefont
  {Wu}}, \bibinfo {author} {\bibfnamefont {Xing-Can}\ \bibnamefont {Yao}},
  \bibinfo {author} {\bibfnamefont {Hao-Ze}\ \bibnamefont {Chen}}, \bibinfo
  {author} {\bibfnamefont {Xiang-Pei}\ \bibnamefont {Liu}}, \bibinfo {author}
  {\bibfnamefont {Xiao-Qiong}\ \bibnamefont {Wang}}, \bibinfo {author}
  {\bibfnamefont {Yu-Ao}\ \bibnamefont {Chen}}, \ and\ \bibinfo {author}
  {\bibfnamefont {Jian-Wei}\ \bibnamefont {Pan}},\ }\bibfield  {title}
  {\enquote {\bibinfo {title} {A quantum degenerate bose--fermi mixture of 41 k
  and 6 li},}\ }\href {http://stacks.iop.org/0953-4075/50/i=9/a=094001}
  {\bibfield  {journal} {\bibinfo  {journal} {Journal of Physics B: Atomic,
  Molecular and Optical Physics}\ }\textbf {\bibinfo {volume} {50}},\ \bibinfo
  {pages} {094001} (\bibinfo {year} {2017})}\BibitemShut {NoStop}%
\bibitem [{\citenamefont {Park}\ \emph {et~al.}(2015)\citenamefont {Park},
  \citenamefont {Will},\ and\ \citenamefont {Zwierlein}}]{Park2015}%
  \BibitemOpen
  \bibfield  {author} {\bibinfo {author} {\bibfnamefont {Jee~Woo}\ \bibnamefont
  {Park}}, \bibinfo {author} {\bibfnamefont {Sebastian~A.}\ \bibnamefont
  {Will}}, \ and\ \bibinfo {author} {\bibfnamefont {Martin~W.}\ \bibnamefont
  {Zwierlein}},\ }\bibfield  {title} {\enquote {\bibinfo {title} {Ultracold
  dipolar gas of fermionic $^{23}\mathrm{Na}^{40}\mathrm{K}$ molecules in their
  absolute ground state},}\ }\href {\doibase 10.1103/PhysRevLett.114.205302}
  {\bibfield  {journal} {\bibinfo  {journal} {Phys. Rev. Lett.}\ }\textbf
  {\bibinfo {volume} {114}},\ \bibinfo {pages} {205302} (\bibinfo {year}
  {2015})}\BibitemShut {NoStop}%
\bibitem [{\citenamefont {Schulze}\ \emph {et~al.}(2018)\citenamefont
  {Schulze}, \citenamefont {Hartmann}, \citenamefont {Voges}, \citenamefont
  {Gempel}, \citenamefont {Tiemann}, \citenamefont {Zenesini},\ and\
  \citenamefont {Ospelkaus}}]{Schulze2018}%
  \BibitemOpen
  \bibfield  {author} {\bibinfo {author} {\bibfnamefont {Torben~A.}\
  \bibnamefont {Schulze}}, \bibinfo {author} {\bibfnamefont {Torsten}\
  \bibnamefont {Hartmann}}, \bibinfo {author} {\bibfnamefont {Kai~K.}\
  \bibnamefont {Voges}}, \bibinfo {author} {\bibfnamefont {Matthias~W.}\
  \bibnamefont {Gempel}}, \bibinfo {author} {\bibfnamefont {Eberhard}\
  \bibnamefont {Tiemann}}, \bibinfo {author} {\bibfnamefont {Alessandro}\
  \bibnamefont {Zenesini}}, \ and\ \bibinfo {author} {\bibfnamefont {Silke}\
  \bibnamefont {Ospelkaus}},\ }\bibfield  {title} {\enquote {\bibinfo {title}
  {Feshbach spectroscopy and dual-species bose-einstein condensation of
  $^{23}\mathrm{Na}\text{\ensuremath{-}}^{39}\mathrm{K}$ mixtures},}\ }\href
  {\doibase 10.1103/PhysRevA.97.023623} {\bibfield  {journal} {\bibinfo
  {journal} {Phys. Rev. A}\ }\textbf {\bibinfo {volume} {97}},\ \bibinfo
  {pages} {023623} (\bibinfo {year} {2018})}\BibitemShut {NoStop}%
\bibitem [{\citenamefont {Heo}\ \emph {et~al.}(2012)\citenamefont {Heo},
  \citenamefont {Wang}, \citenamefont {Christensen}, \citenamefont {Rvachov},
  \citenamefont {Cotta}, \citenamefont {Choi}, \citenamefont {Lee},\ and\
  \citenamefont {Ketterle}}]{Heo2012}%
  \BibitemOpen
  \bibfield  {author} {\bibinfo {author} {\bibfnamefont {Myoung-Sun}\
  \bibnamefont {Heo}}, \bibinfo {author} {\bibfnamefont {Tout~T.}\ \bibnamefont
  {Wang}}, \bibinfo {author} {\bibfnamefont {Caleb~A.}\ \bibnamefont
  {Christensen}}, \bibinfo {author} {\bibfnamefont {Timur~M.}\ \bibnamefont
  {Rvachov}}, \bibinfo {author} {\bibfnamefont {Dylan~A.}\ \bibnamefont
  {Cotta}}, \bibinfo {author} {\bibfnamefont {Jae-Hoon}\ \bibnamefont {Choi}},
  \bibinfo {author} {\bibfnamefont {Ye-Ryoung}\ \bibnamefont {Lee}}, \ and\
  \bibinfo {author} {\bibfnamefont {Wolfgang}\ \bibnamefont {Ketterle}},\
  }\bibfield  {title} {\enquote {\bibinfo {title} {Formation of ultracold
  fermionic nali feshbach molecules},}\ }\href {\doibase
  10.1103/PhysRevA.86.021602} {\bibfield  {journal} {\bibinfo  {journal} {Phys.
  Rev. A}\ }\textbf {\bibinfo {volume} {86}},\ \bibinfo {pages} {021602}
  (\bibinfo {year} {2012})}\BibitemShut {NoStop}%
\bibitem [{\citenamefont {Camacho-Guardian}\ \emph {et~al.}(2018)\citenamefont
  {Camacho-Guardian}, \citenamefont {Pe\~na Ardila}, \citenamefont {Pohl},\
  and\ \citenamefont {Bruun}}]{Camacho2018}%
  \BibitemOpen
  \bibfield  {author} {\bibinfo {author} {\bibfnamefont {A.}~\bibnamefont
  {Camacho-Guardian}}, \bibinfo {author} {\bibfnamefont {L.~A.}\ \bibnamefont
  {Pe\~na Ardila}}, \bibinfo {author} {\bibfnamefont {T.}~\bibnamefont {Pohl}},
  \ and\ \bibinfo {author} {\bibfnamefont {G.~M.}\ \bibnamefont {Bruun}},\
  }\bibfield  {title} {\enquote {\bibinfo {title} {Bipolarons in a
  bose-einstein condensate},}\ }\href {\doibase 10.1103/PhysRevLett.121.013401}
  {\bibfield  {journal} {\bibinfo  {journal} {Phys. Rev. Lett.}\ }\textbf
  {\bibinfo {volume} {121}},\ \bibinfo {pages} {013401} (\bibinfo {year}
  {2018})}\BibitemShut {NoStop}%
\bibitem [{\citenamefont {Devreese}\ and\ \citenamefont
  {Alexandrov}(2009)}]{Devreese2009}%
  \BibitemOpen
  \bibfield  {author} {\bibinfo {author} {\bibfnamefont {Jozef~T}\ \bibnamefont
  {Devreese}}\ and\ \bibinfo {author} {\bibfnamefont {Alexandre~S}\
  \bibnamefont {Alexandrov}},\ }\bibfield  {title} {\enquote {\bibinfo {title}
  {Fr\"ohlich polaron and bipolaron: recent developments},}\ }\href
  {http://stacks.iop.org/0034-4885/72/i=6/a=066501} {\bibfield  {journal}
  {\bibinfo  {journal} {Reports on Progress in Physics}\ }\textbf {\bibinfo
  {volume} {72}},\ \bibinfo {pages} {066501} (\bibinfo {year}
  {2009})}\BibitemShut {NoStop}%
\bibitem [{\citenamefont {Heiselberg}\ \emph {et~al.}(2000)\citenamefont
  {Heiselberg}, \citenamefont {Pethick}, \citenamefont {Smith},\ and\
  \citenamefont {Viverit}}]{Heiselberg2000}%
  \BibitemOpen
  \bibfield  {author} {\bibinfo {author} {\bibfnamefont {H.}~\bibnamefont
  {Heiselberg}}, \bibinfo {author} {\bibfnamefont {C.~J.}\ \bibnamefont
  {Pethick}}, \bibinfo {author} {\bibfnamefont {H.}~\bibnamefont {Smith}}, \
  and\ \bibinfo {author} {\bibfnamefont {L.}~\bibnamefont {Viverit}},\
  }\bibfield  {title} {\enquote {\bibinfo {title} {Influence of induced
  interactions on the superfluid transition in dilute fermi gases},}\ }\href
  {\doibase 10.1103/PhysRevLett.85.2418} {\bibfield  {journal} {\bibinfo
  {journal} {Phys. Rev. Lett.}\ }\textbf {\bibinfo {volume} {85}},\ \bibinfo
  {pages} {2418--2421} (\bibinfo {year} {2000})}\BibitemShut {NoStop}%
\bibitem [{\citenamefont {Efremov}\ and\ \citenamefont
  {Viverit}(2002)}]{Efremov2002}%
  \BibitemOpen
  \bibfield  {author} {\bibinfo {author} {\bibfnamefont {D.~V.}\ \bibnamefont
  {Efremov}}\ and\ \bibinfo {author} {\bibfnamefont {L.}~\bibnamefont
  {Viverit}},\ }\bibfield  {title} {\enquote {\bibinfo {title} {$p$},}\ }\href
  {\doibase 10.1103/PhysRevB.65.134519} {\bibfield  {journal} {\bibinfo
  {journal} {Phys. Rev. B}\ }\textbf {\bibinfo {volume} {65}},\ \bibinfo
  {pages} {134519} (\bibinfo {year} {2002})}\BibitemShut {NoStop}%
\bibitem [{\citenamefont {Illuminati}\ and\ \citenamefont
  {Albus}(2004)}]{Illuminati2004}%
  \BibitemOpen
  \bibfield  {author} {\bibinfo {author} {\bibfnamefont {Fabrizio}\
  \bibnamefont {Illuminati}}\ and\ \bibinfo {author} {\bibfnamefont
  {Alexander}\ \bibnamefont {Albus}},\ }\bibfield  {title} {\enquote {\bibinfo
  {title} {High-temperature atomic superfluidity in lattice bose-fermi
  mixtures},}\ }\href {\doibase 10.1103/PhysRevLett.93.090406} {\bibfield
  {journal} {\bibinfo  {journal} {Phys. Rev. Lett.}\ }\textbf {\bibinfo
  {volume} {93}},\ \bibinfo {pages} {090406} (\bibinfo {year}
  {2004})}\BibitemShut {NoStop}%
\bibitem [{\citenamefont {Suzuki}\ \emph {et~al.}(2008)\citenamefont {Suzuki},
  \citenamefont {Miyakawa},\ and\ \citenamefont {Suzuki}}]{Suzuki2008}%
  \BibitemOpen
  \bibfield  {author} {\bibinfo {author} {\bibfnamefont {Kazunori}\
  \bibnamefont {Suzuki}}, \bibinfo {author} {\bibfnamefont {Takahiko}\
  \bibnamefont {Miyakawa}}, \ and\ \bibinfo {author} {\bibfnamefont {Toru}\
  \bibnamefont {Suzuki}},\ }\bibfield  {title} {\enquote {\bibinfo {title}
  {$p$},}\ }\href {\doibase 10.1103/PhysRevA.77.043629} {\bibfield  {journal}
  {\bibinfo  {journal} {Phys. Rev. A}\ }\textbf {\bibinfo {volume} {77}},\
  \bibinfo {pages} {043629} (\bibinfo {year} {2008})}\BibitemShut {NoStop}%
\bibitem [{\citenamefont {Enss}\ and\ \citenamefont
  {Zwerger}(2009)}]{Enss2009}%
  \BibitemOpen
  \bibfield  {author} {\bibinfo {author} {\bibfnamefont {T.}~\bibnamefont
  {Enss}}\ and\ \bibinfo {author} {\bibfnamefont {W.}~\bibnamefont {Zwerger}},\
  }\bibfield  {title} {\enquote {\bibinfo {title} {Superfluidity near phase
  separation in bose-fermi mixtures},}\ }\href {\doibase
  10.1140/epjb/e2009-00005-y} {\bibfield  {journal} {\bibinfo  {journal} {The
  European Physical Journal B}\ }\textbf {\bibinfo {volume} {68}},\ \bibinfo
  {pages} {383--389} (\bibinfo {year} {2009})}\BibitemShut {NoStop}%
\bibitem [{\citenamefont {Wu}\ and\ \citenamefont {Bruun}(2016)}]{Wu2016}%
  \BibitemOpen
  \bibfield  {author} {\bibinfo {author} {\bibfnamefont {Zhigang}\ \bibnamefont
  {Wu}}\ and\ \bibinfo {author} {\bibfnamefont {G.~M.}\ \bibnamefont {Bruun}},\
  }\bibfield  {title} {\enquote {\bibinfo {title} {Topological superfluid in a
  fermi-bose mixture with a high critical temperature},}\ }\href {\doibase
  10.1103/PhysRevLett.117.245302} {\bibfield  {journal} {\bibinfo  {journal}
  {Phys. Rev. Lett.}\ }\textbf {\bibinfo {volume} {117}},\ \bibinfo {pages}
  {245302} (\bibinfo {year} {2016})}\BibitemShut {NoStop}%
\end{thebibliography}%

\end{document}